\documentclass[aps, prl, twocolumn, amsmath, amssymb, reprint, floatfix]{revtex4-2}

\usepackage{graphicx}
\usepackage{dcolumn}
\usepackage{bm}

\usepackage[colorlinks=true,linkcolor=blue]{hyperref}

\usepackage[utf8]{inputenc}
\usepackage[T1]{fontenc}
\usepackage{mathptmx}

\begin{document}

\preprint{AIP/123-QED}

\title{Texture fluctuations and emergent dynamics in coupled nanomagnets}

\author{Samuel D. Sl\"oetjes}

\author{Bj\"orgvin Hj\"orvarsson}

\author{Vassilios Kapaklis}%

\affiliation{Department of Physics and Astronomy, Uppsala University, Box 516, SE-75120 Uppsala, Sweden}

\begin{abstract}

We analyse the thermal fluctuations of magnetization textures in two stray field coupled elements, forming mesospins. To this end, the energy landscape associated with the thermal dynamics of the textures is mapped out and asymmetric energy barriers are identified. These barriers are modified by changing the gap that separates the mesospins. Moreover, the coupling between the edges leads to an anisotropy in the curvature of the energy surface at the metastable minima. This yields a dynamic mode splitting of the edge modes and affects the attempt switching frequencies. Thus, we elucidate the mechanism with which the magnons in the thermal bath generate the stochastic fluctuations of the magnetization at the edges.
\end{abstract}

\maketitle

Single domain magnetic nanoislands with binary magnetization states - \textit{mesospins} - are used as building blocks for magnetic metamaterials. Contemporary examples are \textit{e.g.} artificial spin ices (ASI) \cite{Wang2006, Gilbert_Shakti, Gilbert_tetris, Perrin_Nature_2016, Ostman_natphys_2018}, Ising chains and lattices \cite{Arnalds_2DIsing, Ostman_Ising_2018}, and reconfigurable magnonic crystals \cite{gliga2020dynamics, gartside2021reconfigurable,kaffash2021tailoring}. These magnetic metamaterials can be viewed as having thermal fluctuations, associated with switching of the magnetic states of the mesospins \cite{kapaklis2014thermal, Andersson2016, Pohlit_susc_2020, Goryca_PRX_2021, Goryca_PRB_2022}, 
in an activated process \cite{koraltan2020dependence, Andersson2016, Pohlit_susc_2020, Skovdal_arXiv_2022}. However, the magnetic fluctuations are not solely restricted to the switching of rigid mesospins, but can also occur in the interior \textit{textures} of the mesospin, which necessitates a more nuanced view of their thermal excitations. Furthermore, a finite temperature leads to the excitation of magnons, which take place on timescales of the magnetization dynamics, much smaller than those of the fluctuations of individual mesospins \cite{bloch1930theorie}. 

Resonances can occur in mesospins when the wavelength of magnons matches their extension. At frequencies below the uniform (Kittel) mode, there is a band gap, characterized by the absence of magnon modes in the interior of the mesospins. However, dynamic edge modes may occur at frequencies inside to the band gap, which can be used for spectral detection of topological defects in ASI \cite{gliga2013spectral}. We have previously confirmed the existence of additional modes \cite{sloetjes_APL_2021, Skovdal_arXiv_2022}, associated with a binary stochastic fluctuation of the magnetization at the edges, from hereon called edge fluctuations. These edge fluctuations cause switching between S- or C-states \cite{Nanny_arXiv_2022}, and their presence results in a residual entropy in magnetic metamaterials \cite{Gliga_PRB_2015, Skovdal_arXiv_2022}. They have recently also been confirmed experimentally, and were found to occur at shorter time- and length-scales than conventional mesospin switching \cite{Skovdal_arXiv_2022}. 
Thus so far, the intra-nanomagnet excitations are well understood, however an understanding of the coupling between intra- and inter-mesospin degrees of freedom has not been developed for magnetic metamaterials.

\begin{figure}
    \centering
   \includegraphics[width=\columnwidth]{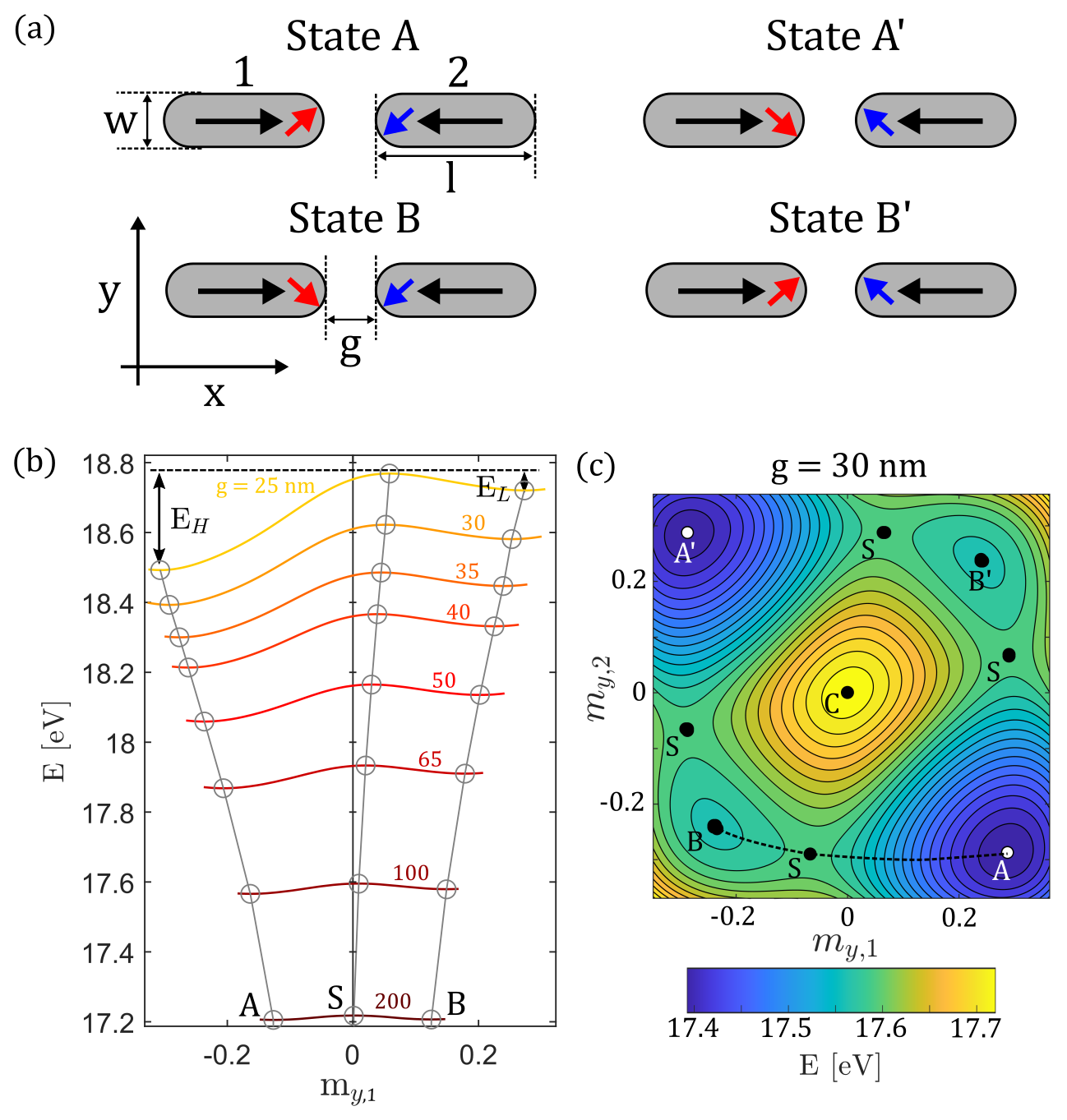}
    \caption{(a) Definition of the the geometrical parameters and the edge magnetization states ($A$ and $B$). (b) The energy barriers associated with switching the edge magnetization of one of the edges, with indication of the points $A$, $S$ and $B$, for different gap values $g$. (c) Morphology of the energy landscape as a function of the perpendicular edge components for a $g=30$ nm gap between the elements. 
   }
    \label{fig:EnergyLandscape}
\end{figure}

For Ising-like mesospins, the edge fluctuations of the texture occur in the transverse direction with respect to the switching of the net magnetization (e.g. in the example of Fig. \ref{fig:EnergyLandscape}a,  $m_y$ is the relevant reaction coordinate in phase space). The height of the barrier associated with these fluctuations depends on the size of the nanomagnets, and the corresponding balance between exchange and demagnetization energies. In the case of a single mesospin, the relevant phase space for the edge fluctuations can be captured in a 1-dimensional energy landscape. When adding one more mesospin, the texture excitations are influences by inter-mesospin interactions, with the phase space extending over a 2-dimensional energy landscape. 
The two nanomagnet system is in the ground state when the two mesospins are magnetized in the same direction, in which case the flux exiting one mesospin will be absorbed by the other, which results in a reduction of the transverse magnetization at the edges. In terms of the energy landscape, this situation leads to a single, global energy minimum, which is associated with to the two edges being aligned. In this case, the coupling causes the splitting of dynamic edge modes into one in-phase and one out-of-phase mode. On the other hand, when the two mesospins are oppositely magnetized (as illustrated in Fig.\ref{fig:EnergyLandscape}a), there will be two degenerate metastable states for the edge magnetization, leading to non-trivial behaviour upon thermal excitation. 

In this work, we discuss the fluctuations of two oppositely magnetized mesospins. To this end, we start by exploring and quantifying the energy landscape associated with the inter-mesospin excitations. We then perform micromagnetic simulations with a temperature implementation, in order to inspect the behaviour of the system under thermal excitation, for a range of different temperatures and spacings between the mesospins. The observed switching events are used to infer the energy barriers and attempt frequencies utilizing Arrhenius analysis \cite{hanggi1990reaction}. Finally, the Arrhenius prefactors are discussed, relating them to the dynamic modes of the system and the curvatures in the energy landscape. 

For the numerical investigation, we performed micromagnetic simulations in \textsc{Mumax$^3$} \cite{mumax3}, which is based on the use of a finite difference method to solve the Landau-Lifshitz-Gilbert (LLG) equation of motion. The size of the stadium shaped mesospin used in this work is $l \times w \times h$ = $360 \times 120 \times 4$~nm$^3$, using rectangular cells of size $2.5 \times 2.5 \times 4$ nm$^3$. We use parameters that are relevant for permalloy, i.e. saturation magnetization $M_s = 1\cdot 10^6$~A/m, an exchange stiffness of $A_{ex}$ = $1 \cdot 10^{-12}$~J/m, and a Gilbert damping of $\alpha = 0.001$. More information on the details of the simulations can be found in \citet{sloetjes_APL_2021}. 

Two mesospins are placed in line, separated by a gap, $g$, that defines the edge-to-edge distance, as illustrated in Fig. \ref{fig:EnergyLandscape}a. 
We define the magnetization in the edges as the averaged transverse magnetization over the half circle that constitutes an edge, i.e. $m_{y,1}=\frac{1}{V}\int_{\mathrm{edge_1}}m_y(\mathbf{r},t) d^3\mathbf{r}$ for the edge of mesospin 1, and similarly for mesospin 2, where $V$ is the volume of the mesospin edge. The lowest energy configuration, state $A$, has opposing transverse magnetization components, $m_{y,1}=-m_{y,2}$, and state $B$ has aligned transverse components, $m_{y,1}=m_{y,2}$, see Fig. \ref{fig:EnergyLandscape}a. 

We represent these states in an energy landscape, parameterized by $m_{y,1}$ and $m_{y,2}$, see Fig. \ref{fig:EnergyLandscape}c. This is an approximate representation of the energy landscape, since the Zeeman energy, $E_Z$, is included in the total energy, therefore it only serves to indicate the morphology. Aside from the states $A$ and $B$ (and the degenerate states $A$´ and $B$´) one can observe the saddle points at $S$ and the top of the energy barrier at $C$. As the gap $g$ between the mesospins increases, the points $A$ and $B$ shift towards the same energy, while the points $S$ and $C$ merge having the same height. Bringing the mesospins closer together causes the energy hill to emerge at point $C$, pushing the states $A$ and $B$ outwards, and breaks the two-fold symmetry. Thus, the gap between the mesospins provides a handle to tailor the energy landscape.

When switching from State $A$ to the opposite State $A$´, it is not energetically favorable to take the direct route $A-C-A$´, corresponding to both moments switching simultaneously. Instead, it is more favorable to take the route $A-B-A$´ via the saddlepoints, which corresponds to consecutive switchings of the nanomagnet edges. Due to the symmetry of the energy landscape, it is only necessary to know the minimum energy path $A-S-B$ for an understanding of the fluctuations. 
We map out the energy barrier along the path that is highlighted in Fig. \ref{fig:EnergyLandscape}c (see Appendix A for the method), the resulting morphologies are shown in Fig. \ref{fig:EnergyLandscape}b. As previously noted, the energy landscape becomes increasingly asymmetric as the gap between the mesospins is reduced. The states $A$ and $B$ are simultaneously pushed away from the center. Furthermore, the saddle point moves away from the center position, towards state $B$. From hereon, we refer to the energy difference between A and S as the high energy barrier ($E_H$), and the energy difference between B and S as the low energy barrier ($E_L$). 

\begin{figure}
    \centering
   \includegraphics[width=1\columnwidth]{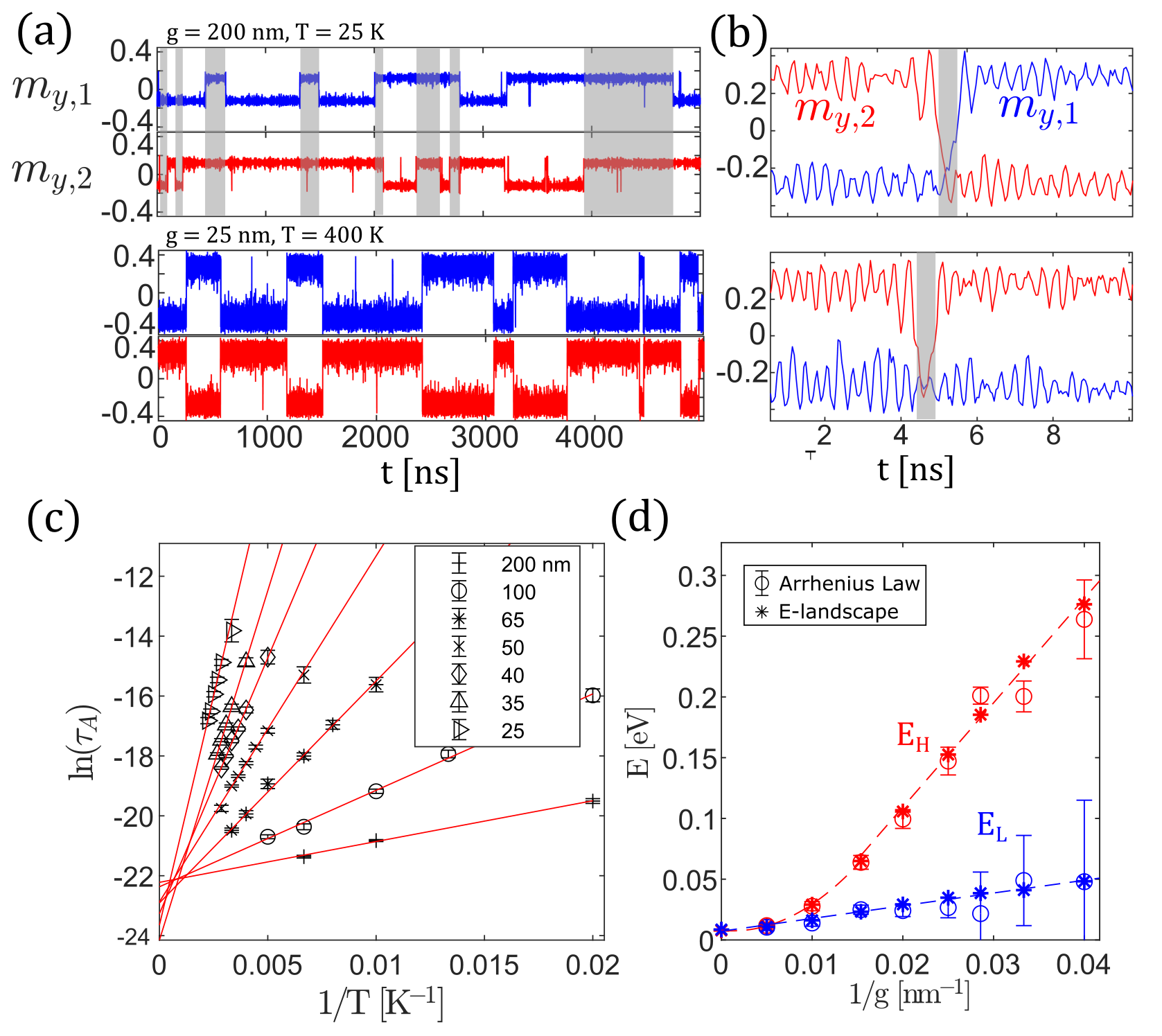}
    \caption{(a) Timetraces of the edge magnetization for two different gaps and temperatures. The shaded regions mark the system being in State $B$, the regions without shading mark state $A$. (b) Switching events for a gap of 25 nm and a temperature of 400 K. Upper panel: a true switching event, lower panel: a false switching event. (c) Arrhenius plots for the relaxation times associated with the high barrier, $\tau_A$. (d) The energy barriers E$_H$ and E$_L$ against the inverse gap values obtained via Arrhenius law, and direct mapping of the energy barrier. The dashed lines are linear fits in the case of $E_L$ and the low gap regime of $E_H$, and a parabolic fit to the high gap regime of $E_H$.
   }
    \label{fig:EnergiesArr}
\end{figure}

Next, we will discuss the thermally induced dynamics of the two oppositely magnetized mesospins. To this end, we use micromagnetic simulations at a finite temperature, as described in  \citet{Leliaert_thermal_mumax3}. The thermal bath is imitated by transforming the LLG equation to a Langevin equation using a stochastic thermal field, given by:
\begin{equation}\label{eq:therm}
    \mu_0\mathbf{H}^{\mathrm{therm}} = \boldsymbol{\eta}\sqrt{\frac{2\alpha k_BT}{M_s\gamma V \Delta t}}
\end{equation}
\noindent
where $k_B$, $T$, $\gamma$, $V$, and $\Delta t$ are the Boltzmann constant, temperature, gyromagnetic ratio, volume of the cell, and the time step, respectively. In addition, $\boldsymbol{\eta}$ is a random vector which changes direction and size at every timestep.
We run the simulations for temperatures from $T = 25$ to 450 K in steps of 50 K, and eight different gaps, ranging from $g=$ 25 to 200~nm.  We analyze the timetraces of the perpendicular component of the magnetization in the two edges $m_{y,j}(t)$, where $j$ denotes the particular edge. Examples of these time traces are shown in Fig. \ref{fig:EnergiesArr}a, for a large ($g=200$~nm) and a small ($g=25$~nm) spacing. Each edge can be seen to switch between the two different states, and the absolute value of the perpendicular magnetization component increases as the spacing becomes smaller. The thermal energy $k_B T$ required to switch the edge magnetization needs to be increased by more than an order of magnitude for $g = 25$~nm compared to $g = 200$~nm, in order to obtain a similar switching rate. Furthermore, the magnetization at the edges is weakly anti-correlated when the mesospins are placed far apart, and strongly anti-correlated when they are close together. This correlation can be quantified using the Pearson correlation coefficient, $\rho (m_{y,1},m_{y,2})$, which is shown in the Appendix C for different spacings and temperatures. 

When the two mesospins are strongly coupled, the edges seemingly switch simultaneously, but zooming in on the switching process (upper panel of Fig. \ref{fig:EnergiesArr}b) reveals that the two edges switch in a consecutive manner, as was predicted on the basis of the energy landscape. Another possibility for a fluctuation is that the edge magnetization switches back to its initial state, as seen in the lower panel of Fig. \ref{fig:EnergiesArr}b. We have observed (see Appendix D) that the process $A-B-A'$ occurs more often that the process $A-B-A$, while in principle these two processes are the same in terms of energy barrier. This asymmetry becomes more pronounced as the gap between the nanomagnets is reduced.

We will now set out to determine the thermal energetics of the textures in the mesospins. The average time in each state is determined, and we use Arrhenius law to extract the energy barriers and attempt frequency:  
\begin{equation}
    \tau_i = \tau_{0,i}e^\frac{E_i}{k_BT}
\end{equation}

where $i$ indicates $A$ or $B$ for $\tau$, and $H$ and $L$ for $E$ respectively, with $\tau_{0,i}$ being the inverse attempt frequency. Generally, for inverse gaps larger than $1/g = 0.027$ nm$^{-1}$, the errors for $E_L$ diverge, due to the high temperatures used for the relatively low barriers, resulting in erroneous estimations of the switching time. For the largest gap, 200 nm, there is no notable difference between E$_L$ and E$_H$, and the energy barrier ($E$~$=$~8 meV) is close to that of a single mesospin \cite{sloetjes_APL_2021} ($E$~$=$~5 meV). Bringing the mesospins closer together leads to an expected increase in $E_H$ and $E_L$. Whereas $E_L$ scales with the inverse gap, $E_H$ shows two different scaling regimes, as shown in Fig. \ref{fig:EnergiesArr}d. It can be seen that up to an inverse gap of $1/g<0.02$ nm$^{-1}$, $E_H$ shows a parabolic scaling with $1/g$, whereas for values $1/g\geq0.02$, there is a linear scaling. The transition between the two regimes occurs between $1/g = 1/65$ nm$^{-1}$ and $1/g = 1/50$ nm$^{-1}$, which means that the radius of the half circle at the edge (60 nm) defines the scale of the transition length. Overall, good agreement can be found between the energies obtained via the stochastic simulations and the energy landscape method, confirming the Arrhenius law. 

\begin{figure}[t!]
    \centering
   \includegraphics[width=1\columnwidth]{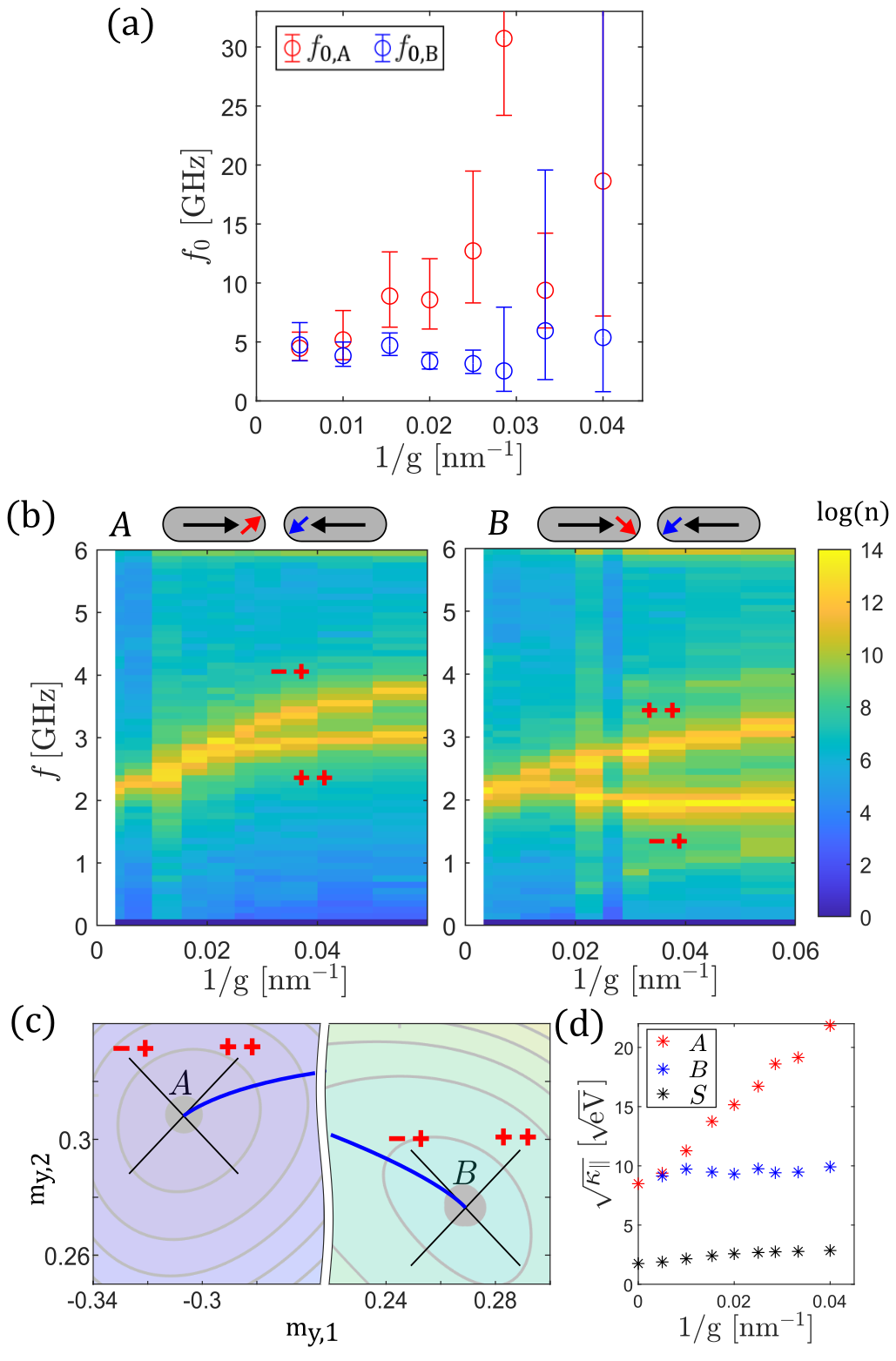}
    \caption{(a) Attempt frequency dependence on the gap for the states $A$ and $B$. (b) Dynamic mode spectra for the states $A$ and $B$, generated by a small thermal field. (c) The minimum energy path emerging from $A$ and $B$ (blue) with the corresponding direction of the in-phase and out-of-phase modes (black). The contours in the background are the same as in Fig. \ref{fig:EnergyLandscape}c (d) Curvatures in the direction of the minimum energy path. 
   }
    \label{fig:edgeScaling}
\end{figure}

Next, we shall focus on the attempt frequency, $f_0$, calculated as the inverse of $\tau_{0,i}$. Similar to the energy barriers, the attempt frequencies for State $A$ and $B$ also show a diverging behaviour upon a decrease of the gap (see Fig. \ref{fig:edgeScaling}a). At the largest spacing, the attempt frequency at $A$ and $B$ become equal, although they are not equal to that of a single mesospin \cite{sloetjes_APL_2021}. Interestingly, it can be observed that $f_{0,A}$ increases when the nanomagnets are being brought closer together, while $f_{0,B}$ shows a slight decrease. In order to understand the scaling of the attempt frequencies, we shall inspect the mode frequencies at the edge of the mesospins, as well as the curvature in the energy landscape at $A$ and $B$. 

The spectrum of magnon modes at the $A$ state is different from those at $B$, see Fig. \ref{fig:edgeScaling}b. The edge modes reside in the frequency band below the Kittel mode ($\sim$~6~GHz) due to the demagnetizing field counteracting the effective field at the edges, thus lowering the eigenfrequencies with respect to the modes inside the mesospin \cite{sloetjes_APL_2021}. We can observe, for both cases, that a mode splitting occurs as the mesospins are brought closer together. The mode splitting in $B$ happens at a larger spacing than the splitting in $A$, which can be understood as being due to the difference in strength of the dipole coupling: in $B$, the sources of the stray field are closer together than in $A$. In both cases, the two resulting branches after the splitting have opposite phases in the in- and out-of-plane directions of the precession. By using phase maps of the mesospins with the full inner texture taken into account (see Appendix E), we determine the in-plane ($\Delta\varphi^{\parallel}$) and out-of-plane ($\Delta\varphi^{\perp}$) phase differences of the modes in the two edges. The out-of-plane components are very small compared to the in-plane ones, i.e. the precessional motion is highly elliptical, therefore we will only consider the in-plane component, where in- and out-of-phase oscillations are indicated by $++$ and $+-$, respectively. From Fig. \ref{fig:edgeScaling}b, it can be seen that the phases of the high and low frequency branches in $B$ are inverted with respect to the branches in $A$. Moreover, at State $A$, both modes increase in frequency as the gap is decreased, whereas in $B$, the $-+$ mode decreases in frequency, which suggests that the $-+$ mode is responsible for switching from $B$ to $A$. 

We turn to the energy landscape to investigate this scenario, where we map the trajectories of the modes at $A$ and $B$, shown in Fig.~\ref{fig:edgeScaling}c, for a gap of $g = 25$ nm. The $-+$ and $++$ modes represent straight paths in the diagonal direction in the $(m_{y,1}, m_{y,2})$ phase space. The mode splitting occurs because the curvatures of the $(m_{y,1},m_{y,2})$-energy landscape at the metastable minima become anisotropic, which is a result of the dipolar coupling between the edges. We plot the minimum energy path from $A$ to $B$ in the vicinity of the minima, and inspect the tangent with the directions of the dynamic modes. At point $A$, the path is mostly tangential to the $++$ mode. In contrast, the path emerging from point $B$ is mostly tangential to the $-+$ mode, which indeed shows that it is the $-+$~mode is the dominant contribution for switching from $B$ to $A$. These findings suggests that the transitions from $A$ to be $B$ can be stimulated using an applied microwave field with a frequency in the $++$ branch, and vice versa with a frequency in the $-+$ branch, although the latter will be problematic in a ferromagnetic resonance experiment, because of the symmetry of the mode. 

Since the damping is low, we can assume that $f_{0,i}$ $\approx$ $f_i$, i.e. the attempt frequency is approximately equal to the eigenmode in the direction of the minimum energy path \cite{kramers1940brownian}. The frequency is related to the curvatures in the energy landscape via $f\propto\sqrt(K)$ \cite{desplat2020entropy}, where $K$ is the total curvature, given in terms of the principle curvatures by $K = \kappa_{\parallel}\kappa_{\perp}$, here $\kappa_j=d^2E/dm_{j}^2$, where $j$ runs over the in- and out-of-plane components (see Appendix B for more details). Thus, upon decreasing $g$, the curvature in the $-+$ direction at point $B$ should decrease, while at point $A$, both curvatures in the $-+$ and $++$ directions increase.
We found that the curvature associated with bringing the magnetization out of the plane, $\kappa_{\perp}$, is more or less constant as a function of $g$, which is due to the strong out-of-plane demagnetizing field overwhelming any effects of proximity to the other edge. The in-plane component, $\kappa_\parallel$ varies strongly, as seen in Fig. \ref{fig:edgeScaling}d. At $A$, $\kappa$ shows an increase with $1/g$, in qualitative agreement with $f_{0,A}$, whereas the curvature at $B$ remains constant, which is not in line with the decrease of $f_{0,B}$. A possible explanation for this discrepancy could be attributed to the path $B-A'$ being favored over $B-A$ as the gap is reduced (see Appendix D), thus suppressing one of the channels for switching. The suppression of this channel effectively lowers the entropic contribution, by reducing the multiplicity. It has been seen before in works by \citet{desplat2018thermal,desplat2020entropy}, that entropic contributions can have large effects on the attempt frequency, which occur when the positive curvatures at the saddle point are small, allowing for an increased number of paths from $A$ to $B$. The positive curvature at the saddle point increases with the height of the energy barrier, thus restricting the saddle point and lowering the amount of paths. However, this effect is compensated by the increase in the curvature at the minima.

In conclusion, we have mapped out the energetics associated with the coupling of intra- and inter-mesospin degrees of freedom in a system of two interacting mesospins. It was found that, upon thermal excitation, a stochastic switching occurs between two degenerate states, $A$ and $B$, and that the reciprocal gap between the nanomagnets provides a handle on the symmetry and barrier height in the energy landscape, modifying in turn the correlations between the edges. 
These correlations feature a strong temperature dependency, that could potentially be harvested in nanomagnetic neuromorphic-like systems, having the opportunity to operate at sub-ps timescales controlled by light, as demonstrated by \citet{Pancaldi:2019gq}, and \citet{femto1}. 
Moreover, by associating the dynamic modes of the system with directions in the energy landscape, we were able to explain the trends in the attempt frequency upon a variation of the gap value. As such, we shed light on how the thermal bath, i.e. the magnonic excitations, couples to the stochastic fluctuations on the next length scale. The stochastic fluctuations of the edges may in turn affect fluctuations on the next length scale, i.e. the full switching of the mesospin. For example, a certain state $A$ can induce a preferred helicity for switching the magnetization in the $m_x$ direction, similar to the symmetry breaking pointed out by \citet{leo2021chiral}. The double degeneracy of the metastable states of the edge magnetization can in this case cause an increase of a factor two in the switching rate of the mesospin island. 
Tunable magnetization texture states have also been recently shown to be useful building block of arrays, where their magnonic properties can be harnessed for hardware implementations of reservoir computing schemes \cite{Reservoir_Japan_2019, ASI_PC_2022}.

\section{Acknowledgments}

We wish to thank Louise Desplat and Matías Pablo Grassi for helpful discussions. The authors acknowledge support STINT (Project No. KO2016-6889) and the Swedish Research Council (Project No. 2019-03581). SDS gratefully acknowledges financial support from the Carl Tryggers Foundation (Project No. 19:175 and 21:1219).

\section*{Author Declarations}

\subsection*{Data availability}
The data that support the findings are available upon request.

\subsection*{Conflict of Interest}

The authors have no conflicts of interest to disclose.

\section{Appendix A: Energy barrier calculations}
\begin{figure}[h!]
    \centering
    \includegraphics[width=0.8\columnwidth]{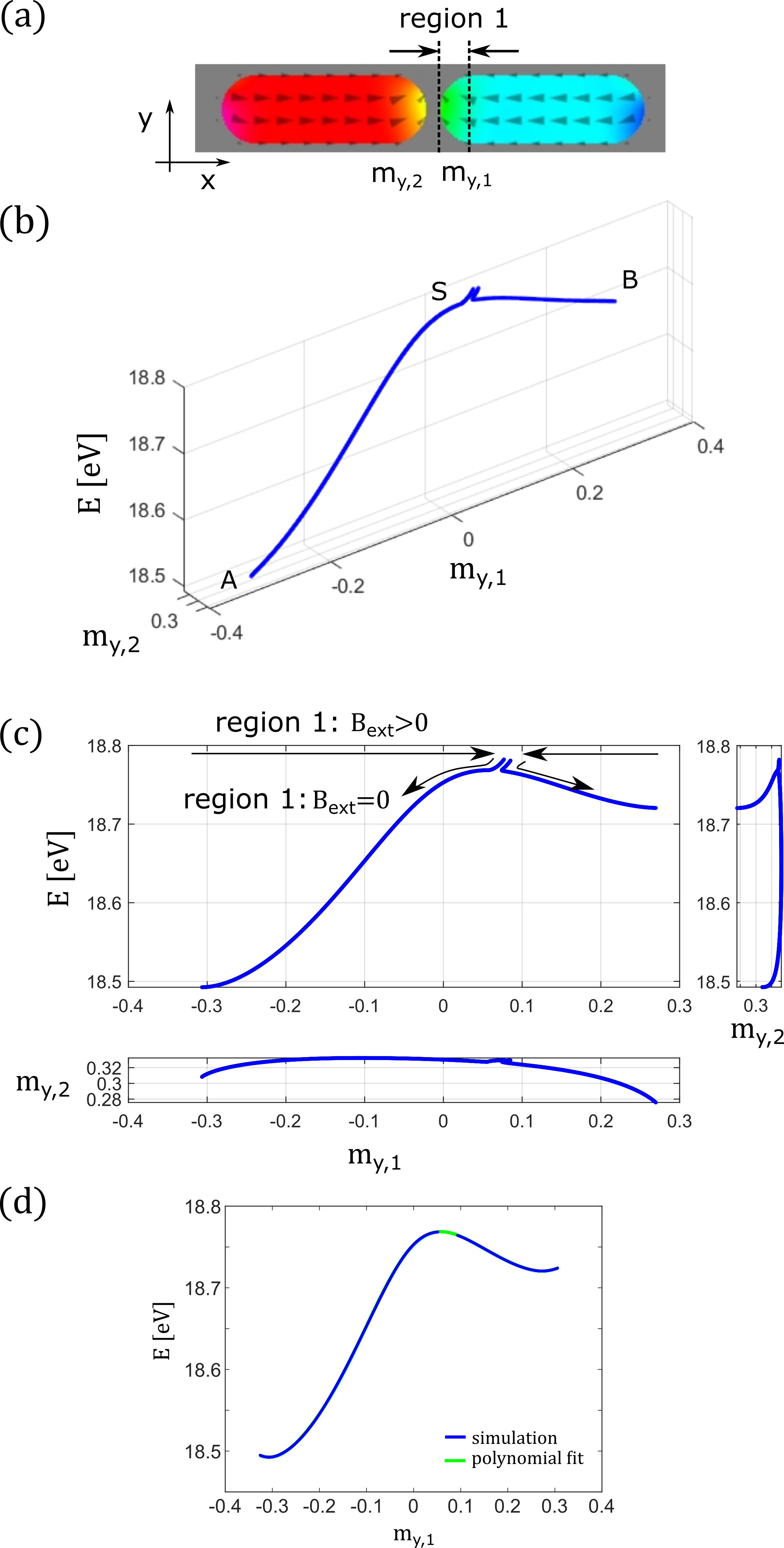}
    \caption{(a) The two mesospins, prepared in state $B$. The field is applied to region 1, which is defined by the half circle that constitutes the edge on mesospin 1. This field pushes the state towards the saddle point $S$. (b) The paths from a point in the vicinity of $S$ to $A$ and to $B$, mapping out the majority of the minimum energy path from $A$ to $B$. (c) The projections of the paths shown in (b) on the $(m_{y,1},E)$, $(m_{y,2},E)$, and the $(m_{y,1},m_{y,2})$-planes. (d) The simulated path with the cusps removed, after fitting to a 10$^{\mathrm{th}}$ order polynomial.}
    \label{fig:EnergyBarrier}
\end{figure}

The energy barrier morphology is mapped out by ``pushing" the system to a point close to the saddle point using an external field, and then letting it relax according to the Landau Lifshitz Gilbert (LLG) equation, which is given by:
\begin{equation}
    \frac{d\mathbf{m}}{dt} = -\frac{\gamma}{1+\alpha^2}[\mathbf{m}\times\mathbf{H}+\alpha\mathbf{m}\times(\mathbf{m}\times\mathbf{H})]
\end{equation}

where $\gamma$, $\alpha$, $\mathbf{m}$, and $\mathbf{H}$ are the gyromagnetic ratio, Gilbert damping constant, magnetization vector, and the effective field, respectively. The effective field acts as a proxy for including the different energy contributions in the equation of motion via
\begin{equation}
    \mathbf{H} = -\frac{1}{\mu_0M_s}\frac{\partial E}{\partial \mathbf{m}}
\end{equation}

where $\mu_0$ is the permeability of free space, and $M_s$ is the saturation magnetization. The total energy is a sum of the separate energies: $E = E_{demag} + E_{exchange} + E_{Zeeman}$. The first term on the right hand side of the LLG equation describes the precession of the moment around the effective field, and the second term describes the movement of the magnetization towards the effective field. For the LLG equation to correctly predict the minimum energy path, the precession needs to be suppressed, therefore we set the damping to the high value of $\alpha=10$. 

The system is first configured in either the $A$ or $B$ state (see Fig. \ref{fig:EnergyBarrier}a). Then, the field is applied to region 1 to push one of the edges to a point close to the saddle point $S$, after which the field is taken away, and the system evolves for 60 ns to either local minimum, $A$ or $B$. During this process, the energy is recorded every $1\times10^{-12}$ s, giving a highly resolved trajectory in phase space. The resulting curves can be seen in Fig. \ref{fig:EnergyBarrier}b,c. The minimum energy path indicates that the majority of the change occurs in $m_{y,1}$, with a small response of $m_{y,2}$.

A cusp is visible near the saddle point, which is a result of the system having relaxed in an external field, thus not being precisely positioned on the minimum energy path. Thus, the cusp signifies the system moving from the position in the field to the minimum energy path. This cusp is relatively small and can be removed, after which we fit a 10$^{\mathrm{th}}$ order polynomial to the curve projected on the $(m_{y,1},E)$-plane. This fitting procedure mostly serves to bridge the gap around the saddle point (see Fig. \ref{fig:EnergyBarrier}d). Strictly, the fitting should take place in three dimensions ($m_{y,1}$ $m_{y,2}$, and $E$), but since $m_{y,2}$ varies very little, we are only interested in the height of the energy barrier and later on the curvature at $S$, fitting the curve only to the projection is justified. With this method we recover the minimum energy path from $A$ to $B$, yielding the energy barriers, $E_L$ and $E_H$. 

This method is relatively simple and can be entirely carried out using \textsc{Mumax3} in the LLG framework. The only drawback is that it demands prior knowledge of the minimum energy path through phase space, in contrast to more sophisticated methods such as the Nudged Elastic Band (NUB) method, and variations thereof \cite{bessarab2015method}.

\section{Appendix B: Curvature calculations}

As stated in the main text, we evaluate the curvatures by taking the second derivative of the energy to the magnetization, $\kappa = \partial^2E/\partial m^2$, at certain points in the energy landscape.

\subsection{Curvature at S}
From the minimum energy path, we obtain the in-plane curvature at the saddle point, $\kappa_{\parallel}^S$, for which $\partial E/\partial m_{y,1}=0$ at the top of the energy barrier. We evaluate the curvature only in the $(m_{y,1},E)$-plane, as the change in the $m_{y,2}$-component is negligible at $S$. 

\begin{figure}[t]
    \centering
    \includegraphics[width=1\columnwidth]{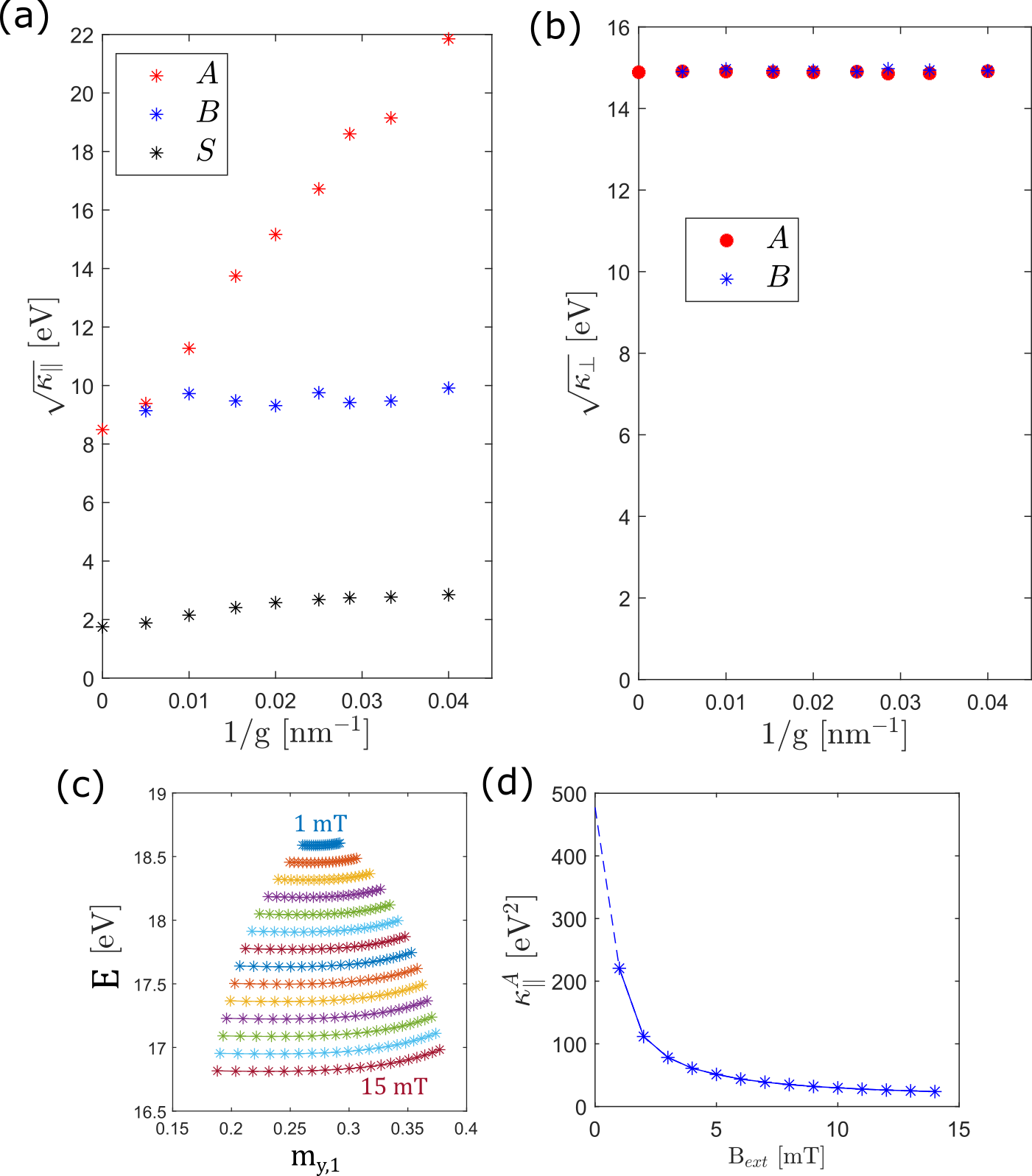}
    \caption{(a) In-plane curvatures of the minimum energy path at $A$, $B$, and $S$. (b) Out of plane curvatures at A and B. (c) The ``wiggling" of the magnetization at $A$, for different applied field strengths, and (d) the curvatures for different guiding fields. The dashed line denotes the extrapolation to 0.}
    \label{fig:Supp_Curvatures}
\end{figure}

\subsection{Curvatures at A and B}
Next we calculate the curvature at $A$ and $B$ via a different method, as the morphology that we previously mapped out only provides one half of the metastable minimum. 
Instead, we use a ``guiding" field parallel to the edge magnetization to wiggle the edge around $A$ and $B$. The guiding field method is similar to the one used for mapping out the energy landscape. However, to obtain the precise value of the curvature, we should eliminate the Zeeman energy, $E_Z$, contribution. In order to eliminate this contribution, we extrapolate to the zero field value of the curvature after calculating the curvature for different sizes of guiding fields, from 15 mT to 1 mT in steps of 1 mT (see Fig. \ref{fig:Supp_Curvatures}c). Mathematically, this can be expressed as:

\begin{equation}
    \kappa^{A,B}_{\parallel} = \lim_{\mathbf{B}_{\mathbf{ext}} \to 0} \frac{\partial^2 E}{\partial m_{y,1}^2}(\mathbf{B}_{\mathbf{ext}})
\end{equation}

The extrapolation method is based on a spline method, performed in \textsc{Matlab}. An example of such an extrapolation for a gap value of $g=25$ nm, is shown in Fig. \ref{fig:Supp_Curvatures}d.

Here, it is necessary to take into account both perpendicular components, $m_{y,1}$ and $m_{y,2}$, since the angle of the path with the $m_{y,1}$ axis can be quite substantial at $A$ and $B$, and varies as a function of 1/g, especially at State $B$ (see Fig. \ref{fig:Supp_Curvatures}d). We can determine the curvature of the parabola projected on the $m_{y,1}$-axis, given by $\partial^2E/\partial m_{y,1}^2$, which in turn provides the total curvature of the parabola through:
\begin{equation}
    E''=\frac{\partial^2E}{\partial m_{y,1}^2}cos^2(\theta)
\end{equation}
\noindent
where $\theta$ is the angle between the $m_{y,1}$ and $m_{y,2}$ components of the path.

The square root of the in-plane curvatures for all the simulated gap values are shown in Fig. \ref{fig:Supp_Curvatures}a. 

\subsection{Out of plane curvatures at A and B}
The out-of-plane curvatures were determined by simply relaxing the magnetization to either $A$ or $B$, by slightly pushing the magnetization out of the plane using an external field and measuring the energy versus the magnetization response, $m_{z,1}$. Here, the contribution from the Zeeman interaction can be neglected, since the field is perpendicular to the magnetization, and $E_{Zeeman}= -\mu_0\mathbf{m}\cdot\mathbf{H}$. A parabola can the be fitted to find the out-of plane curvature, $\kappa^{A,B}_{\perp}$. The result is shown in Fig. \ref{fig:Supp_Curvatures}b.

\section{Appendix C: Correlations}

\begin{figure}[t]
    \centering
    \includegraphics[width=0.8\columnwidth]{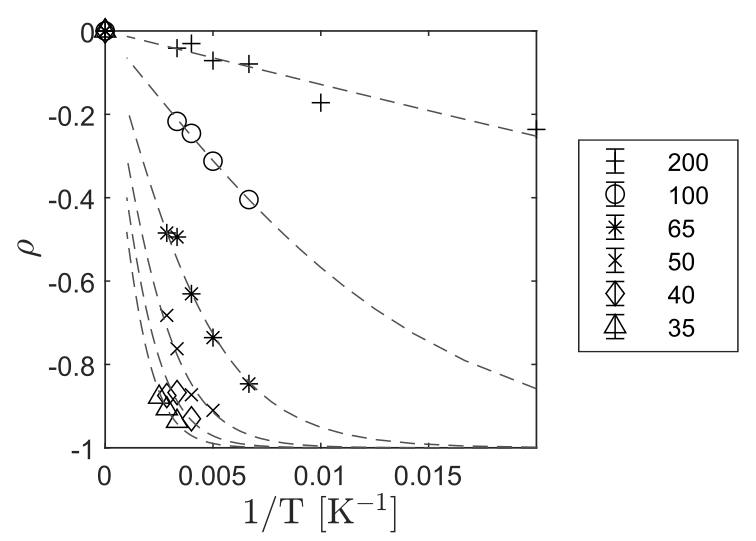}
    \caption{Correlation values: The symbols are calculated from the time traces for the spacings indicated in the legend and in nm. The dashed lines are fits using equation \ref{eq:corrE}.}
    \label{fig:Correlations}
\end{figure}

The edge magnetization correlations were quantified by the Pearson correlation coefficient, given by:

\begin{equation}
    \rho (m_{y,1},m_{y,2}) = \frac{1}{N} \frac{\int(m_{y,1}(t)-\mu_{y,1})(m_{y,2}(t)-\mu_{y,2})dt}{\sigma_{y,1}\sigma_{y,2}}
    \label{eq:corr}
\end{equation}
The results of this calculation for some values of the gap are shown in Fig.~\ref{fig:Correlations}. It can be seen that the correlations decrease with increasing temperature, as well as an increasing gap. We fitted these data points using the following function \footnote{This function can be derived using $\rho (m_{y,1},m_{y,2}) = \frac{1}{T}\int m_{y,1}m_{y,2}dt$, and considering just two intervals, $\tau_A$, and $\tau_B$, in which case $T = \tau_A + \tau_B$. Now, $\tau_A$ runs from $t_0$ to $t_1$, and $\tau_B$ from $t_1$ to $t_2$, thus the integral can be split into the two contributions. Assuming that $m_{y,1}m_{y,2}=-1$ in State A, and $m_{y,1}m_{y,2}=1$ in State B (the normalized case), we obtain $\rho (\tau_A,\tau_B) = \frac{\tau_B - \tau_A}{\tau_A + \tau_B}$, which can be seen to converge to -1 as $\tau_B \rightarrow 0$, and 1 if $\tau_A \rightarrow 0$. Substituting the Arrhenius law, and assuming equal attempt frequencies for A and B gives $\rho = \frac{ 1 - e^{\frac{E_H-E_L}{k_BT}}}{ 1 + e^{\frac{E_H-E_L}{k_BT}}}$, which is equal to Eq. \ref{eq:corrE}.}:

\begin{equation}
     \rho (m_{y,1},m_{y,2}) = -\tanh\left(\frac{E_H - E_L}{2 k_BT}\right)
    \label{eq:corrE}
\end{equation}

This implies that the correlations are only dependent on the difference between the energy barriers, as well as the temperature.

\newpage
\section{Appendix D: Ratio of A-B-A to A-B-A' switching events}

\begin{figure}[t]
    \centering
    \includegraphics[width=0.8\columnwidth]{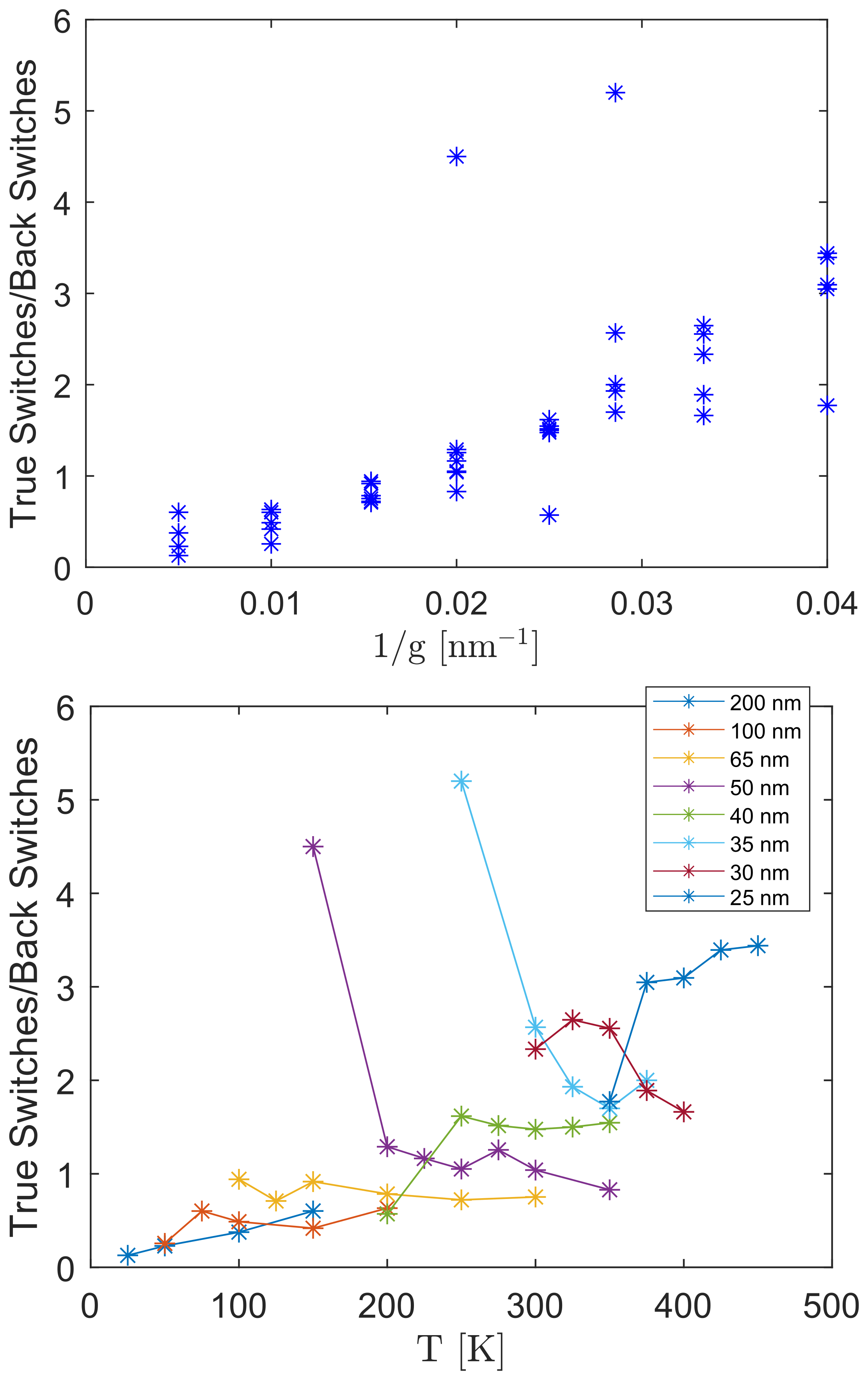}
    \caption{The asymmetry between A-B-A' (`true switches') versus A-B-A (`back switches') against the inverse gap (upper panel) and temperature (lower panel). The two outliers at are likely the result of low statistics.}
    \label{fig:TrueS}
\end{figure}

Figure \ref{fig:TrueS} shows the ratio of switching processes in which the system starts at state A, propagates to state B, and then goes to either state A or A'. It can be seen that this ratio scales mostly with the inverse gap, the temperature not playing a large role. 

\newpage
\section{Appendix E: Dynamic modes and phase determination}

The dynamic modes are produced by applying a thermal field at low temperatures in order to excite all the possible modes. This is done for a given time interval, during which the integrated magnetization at the edges is recorded, and subsequently taking the Fourier transform $\mathbf{m}^{\mathrm{edge}}(f)~=~\mathcal{F}\{\mathbf{m}^{\mathrm{edge}}(t)\}$. The amplitudes are then given by $n^{\mathrm{edge}}(f)~=~|m_{y}^{\mathrm{edge}}(f)|^2 + |m_{z}^{\mathrm{edge}}(f)|^2$. 

The phases of the oscillations were determined by plotting the sine of the phase, multiplied by the magnon intensity. To this end, we first prepare the system in either state A or B using an external magnetic field. We then run the simulation for a certain amount of time, exposed to a low enough temperature such that the system does not switch from A to B or vice versa. During the simulation, we save the spatial magnetization every $5 \cdot 10^{-12}$ s. We then stack these states in a 3D matrix $(x,y,t)$ and take a Fourier transform to the frequency space:
\begin{equation}
    m_{y,z}(x,y,f)=\mathcal{F}\{m_{y,z}(x,y,t)\},
\end{equation}
after which the magnon density can be obtained through:
\begin{equation}
    n(x,y,z,f) = |m_y(x,y,f)|^2 + |m_z(x,y,f)|^2.
\end{equation}

From the Fourier transformation, we can also the obtain the phase information of the magnons by means of the angle of the Fourier components in the complex plane:
\begin{equation}
    \phi_{y,z}(x,y,f) = \angle \mathcal{F}\{m_{y,z}(x,y,t)\}
\end{equation}

\begin{figure}[t]
    \centering
    \includegraphics[width=1\columnwidth]{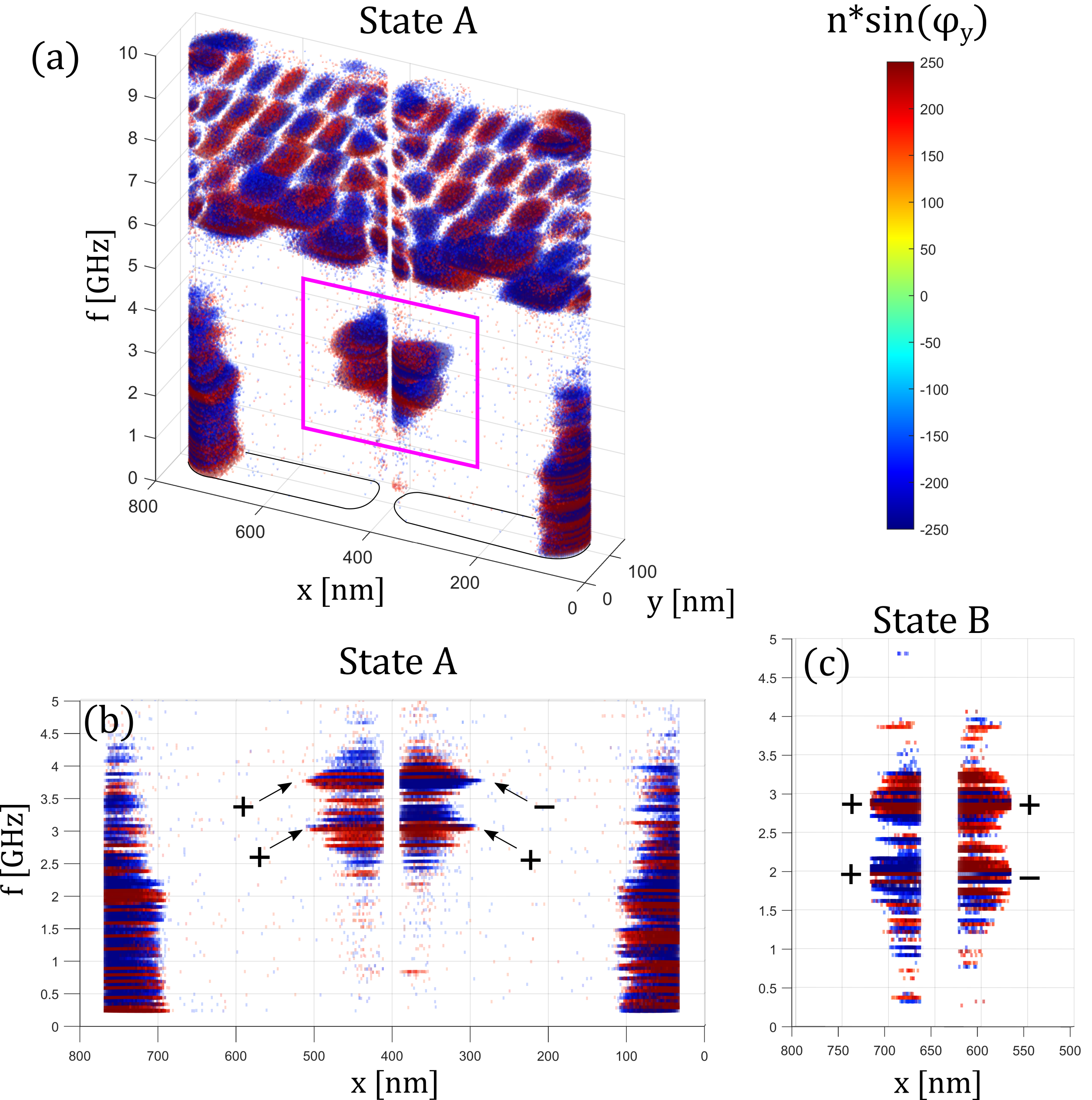}
    \caption{The intensity and phase of the magnonic excitations for (a) State A, in which case the edge excitations lie below 5 GHz, and the inner excitations which lie above this value. The relevant edge excitations are highlighted by a pink square. (b) Side view of the modes, where the two modes representing the in- and out-of-phase modes, can be clearly distinguished. The thermal noise causes some spreading of the mode intensity. (c) The edge mode cross section for State B.}
    \label{fig:Phases}
\end{figure}

We then generate a 3D intensity plot with the phase included as $sin(\phi_y)$, and the amplitude as $n$, giving a total amplitude of $n\times sin(\phi_y)$. The 3D intensity plot is generated using the `\texttt{vol3d.m}' matlab code, which uses voxels to represent intensities in 3D space. The code for \texttt{vol3d.m} is written by \citet{vol3d}, and is available from Matlab File Exchange. The resulting plots are shown in Fig. \ref{fig:Phases}. 

\begin{thebibliography}{36}%
\makeatletter
\providecommand \@ifxundefined [1]{%
 \@ifx{#1\undefined}
}%
\providecommand \@ifnum [1]{%
 \ifnum #1\expandafter \@firstoftwo
 \else \expandafter \@secondoftwo
 \fi
}%
\providecommand \@ifx [1]{%
 \ifx #1\expandafter \@firstoftwo
 \else \expandafter \@secondoftwo
 \fi
}%
\providecommand \natexlab [1]{#1}%
\providecommand \enquote  [1]{``#1''}%
\providecommand \bibnamefont  [1]{#1}%
\providecommand \bibfnamefont [1]{#1}%
\providecommand \citenamefont [1]{#1}%
\providecommand \href@noop [0]{\@secondoftwo}%
\providecommand \href [0]{\begingroup \@sanitize@url \@href}%
\providecommand \@href[1]{\@@startlink{#1}\@@href}%
\providecommand \@@href[1]{\endgroup#1\@@endlink}%
\providecommand \@sanitize@url [0]{\catcode `\\12\catcode `\$12\catcode
  `\&12\catcode `\#12\catcode `\^12\catcode `\_12\catcode `\%12\relax}%
\providecommand \@@startlink[1]{}%
\providecommand \@@endlink[0]{}%
\providecommand \url  [0]{\begingroup\@sanitize@url \@url }%
\providecommand \@url [1]{\endgroup\@href {#1}{\urlprefix }}%
\providecommand \urlprefix  [0]{URL }%
\providecommand \Eprint [0]{\href }%
\providecommand \doibase [0]{https://doi.org/}%
\providecommand \selectlanguage [0]{\@gobble}%
\providecommand \bibinfo  [0]{\@secondoftwo}%
\providecommand \bibfield  [0]{\@secondoftwo}%
\providecommand \translation [1]{[#1]}%
\providecommand \BibitemOpen [0]{}%
\providecommand \bibitemStop [0]{}%
\providecommand \bibitemNoStop [0]{.\EOS\space}%
\providecommand \EOS [0]{\spacefactor3000\relax}%
\providecommand \BibitemShut  [1]{\csname bibitem#1\endcsname}%
\let\auto@bib@innerbib\@empty
\bibitem [{\citenamefont {Wang}\ \emph {et~al.}(2006)\citenamefont {Wang},
  \citenamefont {Nisoli}, \citenamefont {Freitas}, \citenamefont {Li},
  \citenamefont {McConville}, \citenamefont {Cooley}, \citenamefont {Lund},
  \citenamefont {Samarth}, \citenamefont {Leighton}, \citenamefont {Crespi},\
  and\ \citenamefont {Schiffer}}]{Wang2006}%
  \BibitemOpen
  \bibfield  {author} {\bibinfo {author} {\bibfnamefont {R.~F.}\ \bibnamefont
  {Wang}}, \bibinfo {author} {\bibfnamefont {C.}~\bibnamefont {Nisoli}},
  \bibinfo {author} {\bibfnamefont {R.~S.}\ \bibnamefont {Freitas}}, \bibinfo
  {author} {\bibfnamefont {J.}~\bibnamefont {Li}}, \bibinfo {author}
  {\bibfnamefont {W.}~\bibnamefont {McConville}}, \bibinfo {author}
  {\bibfnamefont {B.~J.}\ \bibnamefont {Cooley}}, \bibinfo {author}
  {\bibfnamefont {M.~S.}\ \bibnamefont {Lund}}, \bibinfo {author}
  {\bibfnamefont {N.}~\bibnamefont {Samarth}}, \bibinfo {author} {\bibfnamefont
  {C.}~\bibnamefont {Leighton}}, \bibinfo {author} {\bibfnamefont {V.~H.}\
  \bibnamefont {Crespi}},\ and\ \bibinfo {author} {\bibfnamefont
  {P.}~\bibnamefont {Schiffer}},\ }\href {https://doi.org/10.1038/nature04447}
  {\bibfield  {journal} {\bibinfo  {journal} {Nature}\ }\textbf {\bibinfo
  {volume} {439}},\ \bibinfo {pages} {303} (\bibinfo {year}
  {2006})}\BibitemShut {NoStop}%
\bibitem [{\citenamefont {Gilbert}\ \emph {et~al.}(2014)\citenamefont
  {Gilbert}, \citenamefont {Chern}, \citenamefont {Zhang}, \citenamefont
  {O’Brien}, \citenamefont {Fore}, \citenamefont {Nisoli},\ and\
  \citenamefont {Schiffer}}]{Gilbert_Shakti}%
  \BibitemOpen
  \bibfield  {author} {\bibinfo {author} {\bibfnamefont {I.}~\bibnamefont
  {Gilbert}}, \bibinfo {author} {\bibfnamefont {G.-W.}\ \bibnamefont {Chern}},
  \bibinfo {author} {\bibfnamefont {S.}~\bibnamefont {Zhang}}, \bibinfo
  {author} {\bibfnamefont {L.}~\bibnamefont {O’Brien}}, \bibinfo {author}
  {\bibfnamefont {B.}~\bibnamefont {Fore}}, \bibinfo {author} {\bibfnamefont
  {C.}~\bibnamefont {Nisoli}},\ and\ \bibinfo {author} {\bibfnamefont
  {P.}~\bibnamefont {Schiffer}},\ }\href {https://doi.org/10.1038/nphys3037}
  {\bibfield  {journal} {\bibinfo  {journal} {Nature Physics}\ }\textbf
  {\bibinfo {volume} {10}},\ \bibinfo {pages} {670 } (\bibinfo {year}
  {2014})}\BibitemShut {NoStop}%
\bibitem [{\citenamefont {Gilbert}\ \emph {et~al.}(2016)\citenamefont
  {Gilbert}, \citenamefont {Lao}, \citenamefont {Carrasquillo}, \citenamefont
  {O’Brien}, \citenamefont {Watts}, \citenamefont {Manno}, \citenamefont
  {Leighton}, \citenamefont {Scholl}, \citenamefont {Nisoli},\ and\
  \citenamefont {Schiffer}}]{Gilbert_tetris}%
  \BibitemOpen
  \bibfield  {author} {\bibinfo {author} {\bibfnamefont {I.}~\bibnamefont
  {Gilbert}}, \bibinfo {author} {\bibfnamefont {Y.}~\bibnamefont {Lao}},
  \bibinfo {author} {\bibfnamefont {I.}~\bibnamefont {Carrasquillo}}, \bibinfo
  {author} {\bibfnamefont {L.}~\bibnamefont {O’Brien}}, \bibinfo {author}
  {\bibfnamefont {J.~D.}\ \bibnamefont {Watts}}, \bibinfo {author}
  {\bibfnamefont {M.}~\bibnamefont {Manno}}, \bibinfo {author} {\bibfnamefont
  {C.}~\bibnamefont {Leighton}}, \bibinfo {author} {\bibfnamefont
  {A.}~\bibnamefont {Scholl}}, \bibinfo {author} {\bibfnamefont
  {C.}~\bibnamefont {Nisoli}},\ and\ \bibinfo {author} {\bibfnamefont
  {P.}~\bibnamefont {Schiffer}},\ }\href {https://doi.org/10.1038/nphys3520}
  {\bibfield  {journal} {\bibinfo  {journal} {Nature Physics}\ }\textbf
  {\bibinfo {volume} {12}},\ \bibinfo {pages} {162 } (\bibinfo {year}
  {2016})}\BibitemShut {NoStop}%
\bibitem [{\citenamefont {Perrin}\ \emph {et~al.}(2016)\citenamefont {Perrin},
  \citenamefont {Canals},\ and\ \citenamefont
  {Rougemaille}}]{Perrin_Nature_2016}%
  \BibitemOpen
  \bibfield  {author} {\bibinfo {author} {\bibfnamefont {Y.}~\bibnamefont
  {Perrin}}, \bibinfo {author} {\bibfnamefont {B.}~\bibnamefont {Canals}},\
  and\ \bibinfo {author} {\bibfnamefont {N.}~\bibnamefont {Rougemaille}},\
  }\href {https://doi.org/10.1038/nature20155} {\bibfield  {journal} {\bibinfo
  {journal} {Nature}\ }\textbf {\bibinfo {volume} {540}},\ \bibinfo {pages}
  {410} (\bibinfo {year} {2016})}\BibitemShut {NoStop}%
\bibitem [{\citenamefont {{\"O}stman}\ \emph
  {et~al.}(2018{\natexlab{a}})\citenamefont {{\"O}stman}, \citenamefont
  {Stopfel}, \citenamefont {Chioar}, \citenamefont {Arnalds}, \citenamefont
  {Stein}, \citenamefont {Kapaklis},\ and\ \citenamefont
  {Hj{\"o}rvarsson}}]{Ostman_natphys_2018}%
  \BibitemOpen
  \bibfield  {author} {\bibinfo {author} {\bibfnamefont {E.}~\bibnamefont
  {{\"O}stman}}, \bibinfo {author} {\bibfnamefont {H.}~\bibnamefont {Stopfel}},
  \bibinfo {author} {\bibfnamefont {I.-A.}\ \bibnamefont {Chioar}}, \bibinfo
  {author} {\bibfnamefont {U.~B.}\ \bibnamefont {Arnalds}}, \bibinfo {author}
  {\bibfnamefont {A.}~\bibnamefont {Stein}}, \bibinfo {author} {\bibfnamefont
  {V.}~\bibnamefont {Kapaklis}},\ and\ \bibinfo {author} {\bibfnamefont
  {B.}~\bibnamefont {Hj{\"o}rvarsson}},\ }\href
  {https://doi.org/10.1038/s41567-017-0027-2} {\bibfield  {journal} {\bibinfo
  {journal} {Nature Physics}\ }\textbf {\bibinfo {volume} {14}},\ \bibinfo
  {pages} {375} (\bibinfo {year} {2018}{\natexlab{a}})}\BibitemShut {NoStop}%
\bibitem [{\citenamefont {Arnalds}\ \emph {et~al.}(2016)\citenamefont
  {Arnalds}, \citenamefont {Chico}, \citenamefont {Stopfel}, \citenamefont
  {Kapaklis}, \citenamefont {B{\"a}renbold}, \citenamefont {Verschuuren},
  \citenamefont {Wolff}, \citenamefont {Neu}, \citenamefont {Bergman},\ and\
  \citenamefont {Hj{\"o}rvarsson}}]{Arnalds_2DIsing}%
  \BibitemOpen
  \bibfield  {author} {\bibinfo {author} {\bibfnamefont {U.~B.}\ \bibnamefont
  {Arnalds}}, \bibinfo {author} {\bibfnamefont {J.}~\bibnamefont {Chico}},
  \bibinfo {author} {\bibfnamefont {H.}~\bibnamefont {Stopfel}}, \bibinfo
  {author} {\bibfnamefont {V.}~\bibnamefont {Kapaklis}}, \bibinfo {author}
  {\bibfnamefont {O.}~\bibnamefont {B{\"a}renbold}}, \bibinfo {author}
  {\bibfnamefont {M.~A.}\ \bibnamefont {Verschuuren}}, \bibinfo {author}
  {\bibfnamefont {U.}~\bibnamefont {Wolff}}, \bibinfo {author} {\bibfnamefont
  {V.}~\bibnamefont {Neu}}, \bibinfo {author} {\bibfnamefont {A.}~\bibnamefont
  {Bergman}},\ and\ \bibinfo {author} {\bibfnamefont {B.}~\bibnamefont
  {Hj{\"o}rvarsson}},\ }\href {https://doi.org/10.1088/1367-2630/18/2/023008}
  {\bibfield  {journal} {\bibinfo  {journal} {New Journal of Physics}\ }\textbf
  {\bibinfo {volume} {18}},\ \bibinfo {pages} {023008} (\bibinfo {year}
  {2016})}\BibitemShut {NoStop}%
\bibitem [{\citenamefont {{\"O}stman}\ \emph
  {et~al.}(2018{\natexlab{b}})\citenamefont {{\"O}stman}, \citenamefont
  {Arnalds}, \citenamefont {Kapaklis}, \citenamefont {Taroni},\ and\
  \citenamefont {Hj{\"o}rvarsson}}]{Ostman_Ising_2018}%
  \BibitemOpen
  \bibfield  {author} {\bibinfo {author} {\bibfnamefont {E.}~\bibnamefont
  {{\"O}stman}}, \bibinfo {author} {\bibfnamefont {U.~B.}\ \bibnamefont
  {Arnalds}}, \bibinfo {author} {\bibfnamefont {V.}~\bibnamefont {Kapaklis}},
  \bibinfo {author} {\bibfnamefont {A.}~\bibnamefont {Taroni}},\ and\ \bibinfo
  {author} {\bibfnamefont {B.}~\bibnamefont {Hj{\"o}rvarsson}},\ }\href
  {https://doi.org/10.1088/1361-648X/aad0c1} {\bibfield  {journal} {\bibinfo
  {journal} {Journal of Physics: Condensed Matter}\ }\textbf {\bibinfo {volume}
  {30}},\ \bibinfo {pages} {365301} (\bibinfo {year}
  {2018}{\natexlab{b}})}\BibitemShut {NoStop}%
\bibitem [{\citenamefont {Gliga}\ \emph {et~al.}(2020)\citenamefont {Gliga},
  \citenamefont {Iacocca},\ and\ \citenamefont {Heinonen}}]{gliga2020dynamics}%
  \BibitemOpen
  \bibfield  {author} {\bibinfo {author} {\bibfnamefont {S.}~\bibnamefont
  {Gliga}}, \bibinfo {author} {\bibfnamefont {E.}~\bibnamefont {Iacocca}},\
  and\ \bibinfo {author} {\bibfnamefont {O.~G.}\ \bibnamefont {Heinonen}},\
  }\href {https://doi.org/10.1063/1.5142705} {\bibfield  {journal} {\bibinfo
  {journal} {APL Materials}\ }\textbf {\bibinfo {volume} {8}},\ \bibinfo
  {pages} {040911} (\bibinfo {year} {2020})}\BibitemShut {NoStop}%
\bibitem [{\citenamefont {Gartside}\ \emph {et~al.}(2021)\citenamefont
  {Gartside}, \citenamefont {Vanstone}, \citenamefont {Dion}, \citenamefont
  {Stenning}, \citenamefont {Arroo}, \citenamefont {Kurebayashi},\ and\
  \citenamefont {Branford}}]{gartside2021reconfigurable}%
  \BibitemOpen
  \bibfield  {author} {\bibinfo {author} {\bibfnamefont {J.~C.}\ \bibnamefont
  {Gartside}}, \bibinfo {author} {\bibfnamefont {A.}~\bibnamefont {Vanstone}},
  \bibinfo {author} {\bibfnamefont {T.}~\bibnamefont {Dion}}, \bibinfo {author}
  {\bibfnamefont {K.~D.}\ \bibnamefont {Stenning}}, \bibinfo {author}
  {\bibfnamefont {D.~M.}\ \bibnamefont {Arroo}}, \bibinfo {author}
  {\bibfnamefont {H.}~\bibnamefont {Kurebayashi}},\ and\ \bibinfo {author}
  {\bibfnamefont {W.~R.}\ \bibnamefont {Branford}},\ }\href
  {https://doi.org/10.1038/s41467-021-22723-x} {\bibfield  {journal} {\bibinfo
  {journal} {Nature Communications}\ }\textbf {\bibinfo {volume} {12}},\
  \bibinfo {pages} {2488} (\bibinfo {year} {2021})}\BibitemShut {NoStop}%
\bibitem [{\citenamefont {Kaffash}\ \emph {et~al.}(2021)\citenamefont
  {Kaffash}, \citenamefont {Lendinez},\ and\ \citenamefont
  {Jungfleisch}}]{kaffash2021tailoring}%
  \BibitemOpen
  \bibfield  {author} {\bibinfo {author} {\bibfnamefont {M.~T.}\ \bibnamefont
  {Kaffash}}, \bibinfo {author} {\bibfnamefont {S.}~\bibnamefont {Lendinez}},\
  and\ \bibinfo {author} {\bibfnamefont {M.~B.}\ \bibnamefont {Jungfleisch}},\
  }in\ \href {https://doi.org/10.1109/COMCAS52219.2021.9629059} {\emph
  {\bibinfo {booktitle} {2021 IEEE International Conference on Microwaves,
  Antennas, Communications and Electronic Systems (COMCAS)}}}\ (\bibinfo {year}
  {2021})\ pp.\ \bibinfo {pages} {500--503}\BibitemShut {NoStop}%
\bibitem [{\citenamefont {Kapaklis}\ \emph {et~al.}(2014)\citenamefont
  {Kapaklis}, \citenamefont {Arnalds}, \citenamefont {Farhan}, \citenamefont
  {Chopdekar}, \citenamefont {Balan}, \citenamefont {Scholl}, \citenamefont
  {Heyderman},\ and\ \citenamefont {Hj{\"o}rvarsson}}]{kapaklis2014thermal}%
  \BibitemOpen
  \bibfield  {author} {\bibinfo {author} {\bibfnamefont {V.}~\bibnamefont
  {Kapaklis}}, \bibinfo {author} {\bibfnamefont {U.~B.}\ \bibnamefont
  {Arnalds}}, \bibinfo {author} {\bibfnamefont {A.}~\bibnamefont {Farhan}},
  \bibinfo {author} {\bibfnamefont {R.~V.}\ \bibnamefont {Chopdekar}}, \bibinfo
  {author} {\bibfnamefont {A.}~\bibnamefont {Balan}}, \bibinfo {author}
  {\bibfnamefont {A.}~\bibnamefont {Scholl}}, \bibinfo {author} {\bibfnamefont
  {L.~J.}\ \bibnamefont {Heyderman}},\ and\ \bibinfo {author} {\bibfnamefont
  {B.}~\bibnamefont {Hj{\"o}rvarsson}},\ }\href
  {https://doi.org/10.1038/nnano.2014.104} {\bibfield  {journal} {\bibinfo
  {journal} {Nature nanotechnology}\ }\textbf {\bibinfo {volume} {9}},\
  \bibinfo {pages} {514} (\bibinfo {year} {2014})}\BibitemShut {NoStop}%
\bibitem [{\citenamefont {Andersson}\ \emph {et~al.}(2016)\citenamefont
  {Andersson}, \citenamefont {Pappas}, \citenamefont {Stopfel}, \citenamefont
  {\"Ostman}, \citenamefont {Stein}, \citenamefont {Nordblad}, \citenamefont
  {Mathieu}, \citenamefont {Hj\"orvarsson},\ and\ \citenamefont
  {Kapaklis}}]{Andersson2016}%
  \BibitemOpen
  \bibfield  {author} {\bibinfo {author} {\bibfnamefont {M.~S.}\ \bibnamefont
  {Andersson}}, \bibinfo {author} {\bibfnamefont {S.~D.}\ \bibnamefont
  {Pappas}}, \bibinfo {author} {\bibfnamefont {H.}~\bibnamefont {Stopfel}},
  \bibinfo {author} {\bibfnamefont {E.}~\bibnamefont {\"Ostman}}, \bibinfo
  {author} {\bibfnamefont {A.}~\bibnamefont {Stein}}, \bibinfo {author}
  {\bibfnamefont {P.}~\bibnamefont {Nordblad}}, \bibinfo {author}
  {\bibfnamefont {R.}~\bibnamefont {Mathieu}}, \bibinfo {author} {\bibfnamefont
  {B.}~\bibnamefont {Hj\"orvarsson}},\ and\ \bibinfo {author} {\bibfnamefont
  {V.}~\bibnamefont {Kapaklis}},\ }\href {https://doi.org/10.1038/srep37097}
  {\bibfield  {journal} {\bibinfo  {journal} {Scientific Reports}\ }\textbf
  {\bibinfo {volume} {6}},\ \bibinfo {pages} {37097} (\bibinfo {year}
  {2016})}\BibitemShut {NoStop}%
\bibitem [{\citenamefont {Pohlit}\ \emph {et~al.}(2020)\citenamefont {Pohlit},
  \citenamefont {Muscas}, \citenamefont {Chioar}, \citenamefont {Stopfel},
  \citenamefont {Ciuciulkaite}, \citenamefont {Östman}, \citenamefont
  {Pappas}, \citenamefont {Stein}, \citenamefont {Hjörvarsson}, \citenamefont
  {Jönsson},\ and\ \citenamefont {Kapaklis}}]{Pohlit_susc_2020}%
  \BibitemOpen
  \bibfield  {author} {\bibinfo {author} {\bibfnamefont {M.}~\bibnamefont
  {Pohlit}}, \bibinfo {author} {\bibfnamefont {G.}~\bibnamefont {Muscas}},
  \bibinfo {author} {\bibfnamefont {I.-A.}\ \bibnamefont {Chioar}}, \bibinfo
  {author} {\bibfnamefont {H.}~\bibnamefont {Stopfel}}, \bibinfo {author}
  {\bibfnamefont {A.}~\bibnamefont {Ciuciulkaite}}, \bibinfo {author}
  {\bibfnamefont {E.}~\bibnamefont {Östman}}, \bibinfo {author} {\bibfnamefont
  {S.~D.}\ \bibnamefont {Pappas}}, \bibinfo {author} {\bibfnamefont
  {A.}~\bibnamefont {Stein}}, \bibinfo {author} {\bibfnamefont
  {B.}~\bibnamefont {Hjörvarsson}}, \bibinfo {author} {\bibfnamefont {P.~E.}\
  \bibnamefont {Jönsson}},\ and\ \bibinfo {author} {\bibfnamefont
  {V.}~\bibnamefont {Kapaklis}},\ }\href
  {https://doi.org/10.1103/physrevb.101.134404} {\bibfield  {journal} {\bibinfo
   {journal} {Physical Review B}\ }\textbf {\bibinfo {volume} {101}},\ \bibinfo
  {pages} {134404} (\bibinfo {year} {2020})}\BibitemShut {NoStop}%
\bibitem [{\citenamefont {Goryca}\ \emph {et~al.}(2021)\citenamefont {Goryca},
  \citenamefont {Zhang}, \citenamefont {Li}, \citenamefont {Balk},
  \citenamefont {Watts}, \citenamefont {Leighton}, \citenamefont {Nisoli},
  \citenamefont {Schiffer},\ and\ \citenamefont {Crooker}}]{Goryca_PRX_2021}%
  \BibitemOpen
  \bibfield  {author} {\bibinfo {author} {\bibfnamefont {M.}~\bibnamefont
  {Goryca}}, \bibinfo {author} {\bibfnamefont {X.}~\bibnamefont {Zhang}},
  \bibinfo {author} {\bibfnamefont {J.}~\bibnamefont {Li}}, \bibinfo {author}
  {\bibfnamefont {A.~L.}\ \bibnamefont {Balk}}, \bibinfo {author}
  {\bibfnamefont {J.~D.}\ \bibnamefont {Watts}}, \bibinfo {author}
  {\bibfnamefont {C.}~\bibnamefont {Leighton}}, \bibinfo {author}
  {\bibfnamefont {C.}~\bibnamefont {Nisoli}}, \bibinfo {author} {\bibfnamefont
  {P.}~\bibnamefont {Schiffer}},\ and\ \bibinfo {author} {\bibfnamefont
  {S.~A.}\ \bibnamefont {Crooker}},\ }\href
  {https://doi.org/10.1103/physrevx.11.011042} {\bibfield  {journal} {\bibinfo
  {journal} {Physical Review X}\ }\textbf {\bibinfo {volume} {11}},\ \bibinfo
  {pages} {011042} (\bibinfo {year} {2021})}\BibitemShut {NoStop}%
\bibitem [{\citenamefont {Goryca}\ \emph {et~al.}(2022)\citenamefont {Goryca},
  \citenamefont {Zhang}, \citenamefont {Watts}, \citenamefont {Nisoli},
  \citenamefont {Leighton}, \citenamefont {Schiffer},\ and\ \citenamefont
  {Crooker}}]{Goryca_PRB_2022}%
  \BibitemOpen
  \bibfield  {author} {\bibinfo {author} {\bibfnamefont {M.}~\bibnamefont
  {Goryca}}, \bibinfo {author} {\bibfnamefont {X.}~\bibnamefont {Zhang}},
  \bibinfo {author} {\bibfnamefont {J.~D.}\ \bibnamefont {Watts}}, \bibinfo
  {author} {\bibfnamefont {C.}~\bibnamefont {Nisoli}}, \bibinfo {author}
  {\bibfnamefont {C.}~\bibnamefont {Leighton}}, \bibinfo {author}
  {\bibfnamefont {P.}~\bibnamefont {Schiffer}},\ and\ \bibinfo {author}
  {\bibfnamefont {S.~A.}\ \bibnamefont {Crooker}},\ }\href
  {https://doi.org/10.1103/physrevb.105.094406} {\bibfield  {journal} {\bibinfo
   {journal} {Physical Review B}\ }\textbf {\bibinfo {volume} {105}},\ \bibinfo
  {pages} {094406} (\bibinfo {year} {2022})}\BibitemShut {NoStop}%
\bibitem [{\citenamefont {Koraltan}\ \emph {et~al.}(2020)\citenamefont
  {Koraltan}, \citenamefont {Pancaldi}, \citenamefont {Leo}, \citenamefont
  {Abert}, \citenamefont {Vogler}, \citenamefont {Hofhuis}, \citenamefont
  {Slanovc}, \citenamefont {Bruckner}, \citenamefont {Heistracher},
  \citenamefont {Menniti} \emph {et~al.}}]{koraltan2020dependence}%
  \BibitemOpen
  \bibfield  {author} {\bibinfo {author} {\bibfnamefont {S.}~\bibnamefont
  {Koraltan}}, \bibinfo {author} {\bibfnamefont {M.}~\bibnamefont {Pancaldi}},
  \bibinfo {author} {\bibfnamefont {N.}~\bibnamefont {Leo}}, \bibinfo {author}
  {\bibfnamefont {C.}~\bibnamefont {Abert}}, \bibinfo {author} {\bibfnamefont
  {C.}~\bibnamefont {Vogler}}, \bibinfo {author} {\bibfnamefont
  {K.}~\bibnamefont {Hofhuis}}, \bibinfo {author} {\bibfnamefont
  {F.}~\bibnamefont {Slanovc}}, \bibinfo {author} {\bibfnamefont
  {F.}~\bibnamefont {Bruckner}}, \bibinfo {author} {\bibfnamefont
  {P.}~\bibnamefont {Heistracher}}, \bibinfo {author} {\bibfnamefont
  {M.}~\bibnamefont {Menniti}}, \emph {et~al.},\ }\href
  {https://doi.org/10.1103/PhysRevB.102.064410} {\bibfield  {journal} {\bibinfo
   {journal} {Physical Review B}\ }\textbf {\bibinfo {volume} {102}},\ \bibinfo
  {pages} {064410} (\bibinfo {year} {2020})}\BibitemShut {NoStop}%
\bibitem [{\citenamefont {{Skovdal}}\ \emph {et~al.}(2022)\citenamefont
  {{Skovdal}}, \citenamefont {{Sl{\"o}etjes}}, \citenamefont {{Pohlit}},
  \citenamefont {{Stopfel}}, \citenamefont {{Kapaklis}},\ and\ \citenamefont
  {{Hj{\"o}rvarsson}}}]{Skovdal_arXiv_2022}%
  \BibitemOpen
  \bibfield  {author} {\bibinfo {author} {\bibfnamefont {B.~E.}\ \bibnamefont
  {{Skovdal}}}, \bibinfo {author} {\bibfnamefont {S.~D.}\ \bibnamefont
  {{Sl{\"o}etjes}}}, \bibinfo {author} {\bibfnamefont {M.}~\bibnamefont
  {{Pohlit}}}, \bibinfo {author} {\bibfnamefont {H.}~\bibnamefont {{Stopfel}}},
  \bibinfo {author} {\bibfnamefont {V.}~\bibnamefont {{Kapaklis}}},\ and\
  \bibinfo {author} {\bibfnamefont {B.}~\bibnamefont {{Hj{\"o}rvarsson}}},\
  }\href@noop {} {\bibfield  {journal} {\bibinfo  {journal} {arXiv e-prints}\
  ,\ \bibinfo {eid} {arXiv:2205.00938}} (\bibinfo {year} {2022})},\ \Eprint
  {https://arxiv.org/abs/2205.00938} {arXiv:2205.00938 [cond-mat.mes-hall]}
  \BibitemShut {NoStop}%
\bibitem [{\citenamefont {Bloch}(1930)}]{bloch1930theorie}%
  \BibitemOpen
  \bibfield  {author} {\bibinfo {author} {\bibfnamefont {F.}~\bibnamefont
  {Bloch}},\ }\href@noop {} {\bibfield  {journal} {\bibinfo  {journal}
  {Zeitschrift f{\"u}r Physik}\ }\textbf {\bibinfo {volume} {61}},\ \bibinfo
  {pages} {206} (\bibinfo {year} {1930})}\BibitemShut {NoStop}%
\bibitem [{\citenamefont {Gliga}\ \emph {et~al.}(2013)\citenamefont {Gliga},
  \citenamefont {K{\'a}kay}, \citenamefont {Hertel},\ and\ \citenamefont
  {Heinonen}}]{gliga2013spectral}%
  \BibitemOpen
  \bibfield  {author} {\bibinfo {author} {\bibfnamefont {S.}~\bibnamefont
  {Gliga}}, \bibinfo {author} {\bibfnamefont {A.}~\bibnamefont {K{\'a}kay}},
  \bibinfo {author} {\bibfnamefont {R.}~\bibnamefont {Hertel}},\ and\ \bibinfo
  {author} {\bibfnamefont {O.~G.}\ \bibnamefont {Heinonen}},\ }\href@noop {}
  {\bibfield  {journal} {\bibinfo  {journal} {Physical review letters}\
  }\textbf {\bibinfo {volume} {110}},\ \bibinfo {pages} {117205} (\bibinfo
  {year} {2013})}\BibitemShut {NoStop}%
\bibitem [{\citenamefont {Slöetjes}\ \emph {et~al.}(2021)\citenamefont
  {Slöetjes}, \citenamefont {Hjörvarsson},\ and\ \citenamefont
  {Kapaklis}}]{sloetjes_APL_2021}%
  \BibitemOpen
  \bibfield  {author} {\bibinfo {author} {\bibfnamefont {S.~D.}\ \bibnamefont
  {Slöetjes}}, \bibinfo {author} {\bibfnamefont {B.}~\bibnamefont
  {Hjörvarsson}},\ and\ \bibinfo {author} {\bibfnamefont {V.}~\bibnamefont
  {Kapaklis}},\ }\href {https://doi.org/10.1063/5.0048789} {\bibfield
  {journal} {\bibinfo  {journal} {Applied Physics Letters}\ }\textbf {\bibinfo
  {volume} {118}},\ \bibinfo {pages} {142407} (\bibinfo {year}
  {2021})}\BibitemShut {NoStop}%
\bibitem [{\citenamefont {{Strandqvist}}\ \emph {et~al.}(2022)\citenamefont
  {{Strandqvist}}, \citenamefont {{Skovdal}}, \citenamefont {{Pohlit}},
  \citenamefont {{Stopfel}}, \citenamefont {{Kapaklis}},\ and\ \citenamefont
  {{Hj{\"o}rvarsson}}}]{Nanny_arXiv_2022}%
  \BibitemOpen
  \bibfield  {author} {\bibinfo {author} {\bibfnamefont {N.}~\bibnamefont
  {{Strandqvist}}}, \bibinfo {author} {\bibfnamefont {B.~E.}\ \bibnamefont
  {{Skovdal}}}, \bibinfo {author} {\bibfnamefont {M.}~\bibnamefont {{Pohlit}}},
  \bibinfo {author} {\bibfnamefont {H.}~\bibnamefont {{Stopfel}}}, \bibinfo
  {author} {\bibfnamefont {V.}~\bibnamefont {{Kapaklis}}},\ and\ \bibinfo
  {author} {\bibfnamefont {B.}~\bibnamefont {{Hj{\"o}rvarsson}}},\ }\href@noop
  {} {\bibfield  {journal} {\bibinfo  {journal} {arXiv e-prints}\ ,\ \bibinfo
  {eid} {arXiv:2205.03134}} (\bibinfo {year} {2022})},\ \Eprint
  {https://arxiv.org/abs/2205.03134} {arXiv:2205.03134 [cond-mat.mes-hall]}
  \BibitemShut {NoStop}%
\bibitem [{\citenamefont {Gliga}\ \emph {et~al.}(2015)\citenamefont {Gliga},
  \citenamefont {K{\'a}kay}, \citenamefont {Heyderman}, \citenamefont
  {Hertel},\ and\ \citenamefont {Heinonen}}]{Gliga_PRB_2015}%
  \BibitemOpen
  \bibfield  {author} {\bibinfo {author} {\bibfnamefont {S.}~\bibnamefont
  {Gliga}}, \bibinfo {author} {\bibfnamefont {A.}~\bibnamefont {K{\'a}kay}},
  \bibinfo {author} {\bibfnamefont {L.~J.}\ \bibnamefont {Heyderman}}, \bibinfo
  {author} {\bibfnamefont {R.}~\bibnamefont {Hertel}},\ and\ \bibinfo {author}
  {\bibfnamefont {O.~G.}\ \bibnamefont {Heinonen}},\ }\href
  {https://doi.org/10.1103/PhysRevB.92.060413} {\bibfield  {journal} {\bibinfo
  {journal} {Physical Review B}\ }\textbf {\bibinfo {volume} {92}},\ \bibinfo
  {pages} {060413} (\bibinfo {year} {2015})}\BibitemShut {NoStop}%
\bibitem [{\citenamefont {H{\"a}nggi}\ \emph {et~al.}(1990)\citenamefont
  {H{\"a}nggi}, \citenamefont {Talkner},\ and\ \citenamefont
  {Borkovec}}]{hanggi1990reaction}%
  \BibitemOpen
  \bibfield  {author} {\bibinfo {author} {\bibfnamefont {P.}~\bibnamefont
  {H{\"a}nggi}}, \bibinfo {author} {\bibfnamefont {P.}~\bibnamefont
  {Talkner}},\ and\ \bibinfo {author} {\bibfnamefont {M.}~\bibnamefont
  {Borkovec}},\ }\href {https://doi.org/10.1103/RevModPhys.62.251} {\bibfield
  {journal} {\bibinfo  {journal} {Reviews of Modern Physics}\ }\textbf
  {\bibinfo {volume} {62}},\ \bibinfo {pages} {251} (\bibinfo {year}
  {1990})}\BibitemShut {NoStop}%
\bibitem [{\citenamefont {Vansteenkiste}\ \emph {et~al.}(2014)\citenamefont
  {Vansteenkiste}, \citenamefont {Leliaert}, \citenamefont {Dvornik},
  \citenamefont {Helsen}, \citenamefont {Garcia-Sanchez},\ and\ \citenamefont
  {Van~Waeyenberge}}]{mumax3}%
  \BibitemOpen
  \bibfield  {author} {\bibinfo {author} {\bibfnamefont {A.}~\bibnamefont
  {Vansteenkiste}}, \bibinfo {author} {\bibfnamefont {J.}~\bibnamefont
  {Leliaert}}, \bibinfo {author} {\bibfnamefont {M.}~\bibnamefont {Dvornik}},
  \bibinfo {author} {\bibfnamefont {M.}~\bibnamefont {Helsen}}, \bibinfo
  {author} {\bibfnamefont {F.}~\bibnamefont {Garcia-Sanchez}},\ and\ \bibinfo
  {author} {\bibfnamefont {B.}~\bibnamefont {Van~Waeyenberge}},\ }\href
  {https://doi.org/10.1063/1.4899186} {\bibfield  {journal} {\bibinfo
  {journal} {AIP Advances}\ }\textbf {\bibinfo {volume} {4}},\ \bibinfo {pages}
  {107133} (\bibinfo {year} {2014})}\BibitemShut {NoStop}%
\bibitem [{\citenamefont {Leliaert}\ \emph {et~al.}(2017)\citenamefont
  {Leliaert}, \citenamefont {Mulkers}, \citenamefont {De~Clercq}, \citenamefont
  {Coene}, \citenamefont {Dvornik},\ and\ \citenamefont
  {Van~Waeyenberge}}]{Leliaert_thermal_mumax3}%
  \BibitemOpen
  \bibfield  {author} {\bibinfo {author} {\bibfnamefont {J.}~\bibnamefont
  {Leliaert}}, \bibinfo {author} {\bibfnamefont {J.}~\bibnamefont {Mulkers}},
  \bibinfo {author} {\bibfnamefont {J.}~\bibnamefont {De~Clercq}}, \bibinfo
  {author} {\bibfnamefont {A.}~\bibnamefont {Coene}}, \bibinfo {author}
  {\bibfnamefont {M.}~\bibnamefont {Dvornik}},\ and\ \bibinfo {author}
  {\bibfnamefont {B.}~\bibnamefont {Van~Waeyenberge}},\ }\href
  {https://doi.org/10.1063/1.5003957} {\bibfield  {journal} {\bibinfo
  {journal} {AIP Advances}\ }\textbf {\bibinfo {volume} {7}},\ \bibinfo {pages}
  {125010} (\bibinfo {year} {2017})}\BibitemShut {NoStop}%
\bibitem [{\citenamefont {Kramers}(1940)}]{kramers1940brownian}%
  \BibitemOpen
  \bibfield  {author} {\bibinfo {author} {\bibfnamefont {H.~A.}\ \bibnamefont
  {Kramers}},\ }\href {https://doi.org/10.1016/S0031-8914(40)90098-2}
  {\bibfield  {journal} {\bibinfo  {journal} {Physica}\ }\textbf {\bibinfo
  {volume} {7}},\ \bibinfo {pages} {284} (\bibinfo {year} {1940})}\BibitemShut
  {NoStop}%
\bibitem [{\citenamefont {Desplat}\ and\ \citenamefont
  {Kim}(2020)}]{desplat2020entropy}%
  \BibitemOpen
  \bibfield  {author} {\bibinfo {author} {\bibfnamefont {L.}~\bibnamefont
  {Desplat}}\ and\ \bibinfo {author} {\bibfnamefont {J.-V.}\ \bibnamefont
  {Kim}},\ }\href {https://doi.org/10.1103/PhysRevLett.125.107201} {\bibfield
  {journal} {\bibinfo  {journal} {Physical Review Letters}\ }\textbf {\bibinfo
  {volume} {125}},\ \bibinfo {pages} {107201} (\bibinfo {year}
  {2020})}\BibitemShut {NoStop}%
\bibitem [{\citenamefont {Desplat}\ \emph {et~al.}(2018)\citenamefont
  {Desplat}, \citenamefont {Suess}, \citenamefont {Kim},\ and\ \citenamefont
  {Stamps}}]{desplat2018thermal}%
  \BibitemOpen
  \bibfield  {author} {\bibinfo {author} {\bibfnamefont {L.}~\bibnamefont
  {Desplat}}, \bibinfo {author} {\bibfnamefont {D.}~\bibnamefont {Suess}},
  \bibinfo {author} {\bibfnamefont {J.-V.}\ \bibnamefont {Kim}},\ and\ \bibinfo
  {author} {\bibfnamefont {R.}~\bibnamefont {Stamps}},\ }\href
  {https://doi.org/10.1103/PhysRevB.98.134407} {\bibfield  {journal} {\bibinfo
  {journal} {Physical Review B}\ }\textbf {\bibinfo {volume} {98}},\ \bibinfo
  {pages} {134407} (\bibinfo {year} {2018})}\BibitemShut {NoStop}%
\bibitem [{\citenamefont {Pancaldi}\ \emph {et~al.}(2019)\citenamefont
  {Pancaldi}, \citenamefont {Leo},\ and\ \citenamefont
  {Vavassori}}]{Pancaldi:2019gq}%
  \BibitemOpen
  \bibfield  {author} {\bibinfo {author} {\bibfnamefont {M.}~\bibnamefont
  {Pancaldi}}, \bibinfo {author} {\bibfnamefont {N.}~\bibnamefont {Leo}},\ and\
  \bibinfo {author} {\bibfnamefont {P.}~\bibnamefont {Vavassori}},\ }\href
  {https://doi.org/10.1039/c9nr01628g} {\bibfield  {journal} {\bibinfo
  {journal} {Nanoscale}\ }\textbf {\bibinfo {volume} {11}},\ \bibinfo {pages}
  {7656 } (\bibinfo {year} {2019})}\BibitemShut {NoStop}%
\bibitem [{\citenamefont {Mishra}\ \emph {et~al.}(2021)\citenamefont {Mishra},
  \citenamefont {Ciuciulkaite}, \citenamefont {Zapata-Herrera}, \citenamefont
  {Vavassori}, \citenamefont {Kapaklis}, \citenamefont {Rasing}, \citenamefont
  {Dmitriev}, \citenamefont {Kimel},\ and\ \citenamefont {Kirilyuk}}]{femto1}%
  \BibitemOpen
  \bibfield  {author} {\bibinfo {author} {\bibfnamefont {K.}~\bibnamefont
  {Mishra}}, \bibinfo {author} {\bibfnamefont {A.}~\bibnamefont
  {Ciuciulkaite}}, \bibinfo {author} {\bibfnamefont {M.}~\bibnamefont
  {Zapata-Herrera}}, \bibinfo {author} {\bibfnamefont {P.}~\bibnamefont
  {Vavassori}}, \bibinfo {author} {\bibfnamefont {V.}~\bibnamefont {Kapaklis}},
  \bibinfo {author} {\bibfnamefont {T.}~\bibnamefont {Rasing}}, \bibinfo
  {author} {\bibfnamefont {A.}~\bibnamefont {Dmitriev}}, \bibinfo {author}
  {\bibfnamefont {A.}~\bibnamefont {Kimel}},\ and\ \bibinfo {author}
  {\bibfnamefont {A.}~\bibnamefont {Kirilyuk}},\ }\href
  {https://doi.org/10.1039/D1NR04308K} {\bibfield  {journal} {\bibinfo
  {journal} {Nanoscale}\ }\textbf {\bibinfo {volume} {13}},\ \bibinfo {pages}
  {19367} (\bibinfo {year} {2021})}\BibitemShut {NoStop}%
\bibitem [{\citenamefont {Leo}\ \emph {et~al.}(2021)\citenamefont {Leo},
  \citenamefont {Pancaldi}, \citenamefont {Koraltan}, \citenamefont
  {González}, \citenamefont {Abert}, \citenamefont {Vogler}, \citenamefont
  {Slanovc}, \citenamefont {Bruckner}, \citenamefont {Heistracher},
  \citenamefont {Hofhuis}, \citenamefont {Menniti}, \citenamefont {Suess},\
  and\ \citenamefont {Vavassori}}]{leo2021chiral}%
  \BibitemOpen
  \bibfield  {author} {\bibinfo {author} {\bibfnamefont {N.}~\bibnamefont
  {Leo}}, \bibinfo {author} {\bibfnamefont {M.}~\bibnamefont {Pancaldi}},
  \bibinfo {author} {\bibfnamefont {S.}~\bibnamefont {Koraltan}}, \bibinfo
  {author} {\bibfnamefont {P.~V.}\ \bibnamefont {González}}, \bibinfo {author}
  {\bibfnamefont {C.}~\bibnamefont {Abert}}, \bibinfo {author} {\bibfnamefont
  {C.}~\bibnamefont {Vogler}}, \bibinfo {author} {\bibfnamefont
  {F.}~\bibnamefont {Slanovc}}, \bibinfo {author} {\bibfnamefont
  {F.}~\bibnamefont {Bruckner}}, \bibinfo {author} {\bibfnamefont
  {P.}~\bibnamefont {Heistracher}}, \bibinfo {author} {\bibfnamefont
  {K.}~\bibnamefont {Hofhuis}}, \bibinfo {author} {\bibfnamefont
  {M.}~\bibnamefont {Menniti}}, \bibinfo {author} {\bibfnamefont
  {D.}~\bibnamefont {Suess}},\ and\ \bibinfo {author} {\bibfnamefont
  {P.}~\bibnamefont {Vavassori}},\ }\href
  {https://doi.org/10.1088/1367-2630/abe3ad} {\bibfield  {journal} {\bibinfo
  {journal} {New Journal of Physics}\ }\textbf {\bibinfo {volume} {23}},\
  \bibinfo {pages} {033024} (\bibinfo {year} {2021})},\ \Eprint
  {https://arxiv.org/abs/2010.11291} {2010.11291} \BibitemShut {NoStop}%
\bibitem [{\citenamefont {Nomura}\ \emph {et~al.}(2019)\citenamefont {Nomura},
  \citenamefont {Furuta}, \citenamefont {Tsujimoto}, \citenamefont
  {Kuwabiraki}, \citenamefont {Peper}, \citenamefont {Tamura}, \citenamefont
  {Miwa}, \citenamefont {Goto}, \citenamefont {Nakatani},\ and\ \citenamefont
  {Suzuki}}]{Reservoir_Japan_2019}%
  \BibitemOpen
  \bibfield  {author} {\bibinfo {author} {\bibfnamefont {H.}~\bibnamefont
  {Nomura}}, \bibinfo {author} {\bibfnamefont {T.}~\bibnamefont {Furuta}},
  \bibinfo {author} {\bibfnamefont {K.}~\bibnamefont {Tsujimoto}}, \bibinfo
  {author} {\bibfnamefont {Y.}~\bibnamefont {Kuwabiraki}}, \bibinfo {author}
  {\bibfnamefont {F.}~\bibnamefont {Peper}}, \bibinfo {author} {\bibfnamefont
  {E.}~\bibnamefont {Tamura}}, \bibinfo {author} {\bibfnamefont
  {S.}~\bibnamefont {Miwa}}, \bibinfo {author} {\bibfnamefont {M.}~\bibnamefont
  {Goto}}, \bibinfo {author} {\bibfnamefont {R.}~\bibnamefont {Nakatani}},\
  and\ \bibinfo {author} {\bibfnamefont {Y.}~\bibnamefont {Suzuki}},\ }\href
  {https://doi.org/10.7567/1347-4065/ab2406} {\bibfield  {journal} {\bibinfo
  {journal} {Japanese Journal of Applied Physics}\ }\textbf {\bibinfo {volume}
  {58}},\ \bibinfo {pages} {070901} (\bibinfo {year} {2019})}\BibitemShut
  {NoStop}%
\bibitem [{\citenamefont {Gartside}\ \emph {et~al.}(2022)\citenamefont
  {Gartside}, \citenamefont {Stenning}, \citenamefont {Vanstone}, \citenamefont
  {Holder}, \citenamefont {Arroo}, \citenamefont {Dion}, \citenamefont
  {Caravelli}, \citenamefont {Kurebayashi},\ and\ \citenamefont
  {Branford}}]{ASI_PC_2022}%
  \BibitemOpen
  \bibfield  {author} {\bibinfo {author} {\bibfnamefont {J.~C.}\ \bibnamefont
  {Gartside}}, \bibinfo {author} {\bibfnamefont {K.~D.}\ \bibnamefont
  {Stenning}}, \bibinfo {author} {\bibfnamefont {A.}~\bibnamefont {Vanstone}},
  \bibinfo {author} {\bibfnamefont {H.~H.}\ \bibnamefont {Holder}}, \bibinfo
  {author} {\bibfnamefont {D.~M.}\ \bibnamefont {Arroo}}, \bibinfo {author}
  {\bibfnamefont {T.}~\bibnamefont {Dion}}, \bibinfo {author} {\bibfnamefont
  {F.}~\bibnamefont {Caravelli}}, \bibinfo {author} {\bibfnamefont
  {H.}~\bibnamefont {Kurebayashi}},\ and\ \bibinfo {author} {\bibfnamefont
  {W.~R.}\ \bibnamefont {Branford}},\ }\bibfield  {journal} {\bibinfo
  {journal} {Nature Nanotechnology}\ }\href
  {https://doi.org/10.1038/s41565-022-01091-7} {10.1038/s41565-022-01091-7}
  (\bibinfo {year} {2022})\BibitemShut {NoStop}%
\bibitem [{\citenamefont {Bessarab}\ \emph {et~al.}(2015)\citenamefont
  {Bessarab}, \citenamefont {Uzdin},\ and\ \citenamefont
  {J{\'o}nsson}}]{bessarab2015method}%
  \BibitemOpen
  \bibfield  {author} {\bibinfo {author} {\bibfnamefont {P.~F.}\ \bibnamefont
  {Bessarab}}, \bibinfo {author} {\bibfnamefont {V.~M.}\ \bibnamefont
  {Uzdin}},\ and\ \bibinfo {author} {\bibfnamefont {H.}~\bibnamefont
  {J{\'o}nsson}},\ }\href@noop {} {\bibfield  {journal} {\bibinfo  {journal}
  {Computer Physics Communications}\ }\textbf {\bibinfo {volume} {196}},\
  \bibinfo {pages} {335} (\bibinfo {year} {2015})}\BibitemShut {NoStop}%
\bibitem [{Note1()}]{Note1}%
  \BibitemOpen
  \bibinfo {note} {This function can be derived using $\rho (m_{y,1},m_{y,2}) =
  \protect \frac {1}{T}\DOTSI \intop \ilimits@ m_{y,1}m_{y,2}dt$, and
  considering just two intervals, $\tau _A$, and $\tau _B$, in which case $T =
  \tau _A + \tau _B$. Now, $\tau _A$ runs from $t_0$ to $t_1$, and $\tau _B$
  from $t_1$ to $t_2$, thus the integral can be split into the two
  contributions. Assuming that $m_{y,1}m_{y,2}=-1$ in State A, and
  $m_{y,1}m_{y,2}=1$ in State B (the normalized case), we obtain $\rho (\tau
  _A,\tau _B) = \protect \frac {\tau _B - \tau _A}{\tau _A + \tau _B}$, which
  can be seen to converge to -1 as $\tau _B \rightarrow 0$, and 1 if $\tau _A
  \rightarrow 0$. Substituting the Arrhenius law, and assuming equal attempt
  frequencies for A and B gives $\rho = \protect \frac { 1 - e^{\protect \frac
  {E_H-E_L}{k_BT}}}{ 1 + e^{\protect \frac {E_H-E_L}{k_BT}}}$, which is equal
  to Eq. \ref {eq:corrE}.}\BibitemShut {Stop}%
\bibitem [{\citenamefont {Woodford}(2018)}]{vol3d}%
  \BibitemOpen
  \bibfield  {author} {\bibinfo {author} {\bibfnamefont {O.}~\bibnamefont
  {Woodford}},\ }\href@noop {} {\bibinfo {title} {3-d volume (voxel)
  rendering}},\ \bibinfo {howpublished}
  {\url{https://se.mathworks.com/matlabcentral/fileexchange/22940-vol3d-v/}}
  (\bibinfo {year} {2018}),\ \bibinfo {note} {[Online; accessed 16 May
  2022]}\BibitemShut {NoStop}%
\end{thebibliography}
\end{document}